\documentclass[aps,nofootinbib,groupedaddress]{revtex4}

\usepackage{url}
\usepackage{graphicx}
\usepackage{amsmath,amssymb,amstext,amssymb,amsfonts,amsthm}
\usepackage{hyperref}
\usepackage{appendix}
\usepackage{aas_macros}
\usepackage[margin=1.0in,papersize={8.5in,11in}]{geometry}
\usepackage[utf8]{inputenc}
\usepackage{soul}
\usepackage{color}
\usepackage[normalem]{ulem}

\newcommand{\Wop}[3]{ \boldsymbol{\mathcal{W}}_{#1;#2#3} }

\usepackage{tabularx}
\newcolumntype{C}{>{\centering\arraybackslash}X}
\newcolumntype{R}{>{\raggedleft\arraybackslash}X}

\def\eqref#1{(\ref{#1})}

\begin{document}

\title{Maximum likelihood kinetic Sunyaev Zel'dovich velocity reconstruction}

\author{Dagoberto Contreras${}^{1}$}

\author{Fiona McCarthy${}^{2,3,4}$}

\author{Matthew C. Johnson${}^{1,3}$}

\affiliation{${}^1$Department of Physics and Astronomy, York University, Toronto, Ontario, M3J 1P3, Canada}
\affiliation{${}^2$Center for Computational Astrophysics, Flatiron Institute, 162 5th Avenue, New York, NY 10010 USA}
\affiliation{${}^3$Perimeter Institute for Theoretical Physics, Waterloo, Ontario N2L 2Y5, Canada}
\affiliation{${}^4$Department of Physics and Astronomy, University of Waterloo, Waterloo, Ontario, N2L 3G1, Canada}

\begin{abstract}
The kinetic Sunyaev Zel'dovich (kSZ) effect, cosmic microwave background (CMB) temperature anisotropies induced by the scattering of CMB photons from free electrons, will be measured by near-term CMB experiments at high significance. By combining CMB temperature anisotropies with a tracer of structure, such as a galaxy redshift survey, previous literature introduced a number of techniques to reconstruct the radial velocity field. This reconstructed radial velocity field encapsulates the majority of the cosmological information contained in the kSZ temperature anisotropies, and can provide powerful new tests of the standard cosmological model and theories beyond it. In this paper, we introduce a new estimator for the radial velocity field based on a coarse-grained maximum likelihood fit for the kSZ component of the temperature anisotropies, given a tracer of the optical depth. We demonstrate that this maximum likelihood estimator yields a higher fidelity reconstruction than existing quadratic estimators in the low-noise and high-resolution regime targeted by upcoming CMB experiments. We describe implementations of the maximum likelihood estimator in both harmonic space and map space, using either direct measurements of the optical depth field or a galaxy survey as a tracer. We comment briefly on the impact of biases introduced by imperfect reconstruction of the optical depth field.
\end{abstract}

\maketitle

\section{Introduction}

Future cosmic microwave background (CMB) experiments such as Simons Observatory~\cite{Ade:2018sbj}, CMB-S4~\cite{1610.02743}, and CMB-HD~\cite{Sehgal:2019ewc} will probe the high-resolution, low-noise frontier where secondary CMB anisotropies dominate. Secondary anisotropies arise due to the electromagnetic (Sunyaev Zel'dovich effect) or gravitational (lensing) scattering of CMB photons from structure, and encode a significant amount of astrophysical and cosmological information. How to access this information is an active area of investigation. In this paper, we focus on the kinetic Sunyaev Zel'dovich (kSZ) effect, blackbody temperature anisotropies induced by the scattering of CMB photons from free electrons in bulk motion along the line of sight~\cite{SZ80}. We introduce a new estimator for the radial velocity field based on a coarse-grained maximum likelihood solution constructed from the CMB temperature anisotropies and a tracer of the optical depth. 

Previous literature introduced a set of quadratic estimators (QEs) for the radial velocity field~\cite{Deutsch:2017ybc,Smith:2018bpn,Cayuso:2021ljq} based on the CMB temperature anisotropies and a galaxy survey as a tracer of the optical depth\footnote{More precisely, one reconstructs the remote dipole field: the CMB dipole seen in the local rest frame of the scatterer, projected along the line of sight. The largest contribution to the remote dipole field is given by the radial peculiar velocity, but there are additional 'primordial' contributions; see Refs.~\cite{Terrana2016,Deutsch:2017ybc} for discussion.}. Fisher forecasts~\cite{Deutsch:2017ybc,Smith:2018bpn,Cayuso:2021ljq}, the analysis of N-body simulations~\cite{PhysRevD.98.063502,Giri:2020pkk}, and the analysis of Gaussian mocks containing various foregrounds and systematics \cite{Cayuso:2021ljq} indicate that these QEs will yield high fidelity reconstructions of the radial velocity field with near-term CMB and galaxy surveys. The QE reconstructs the radial velocity field from statistical anisotropies in the cross-power spectrum of the CMB temperature and a redshift-binned galaxy survey. The estimator relies on the assumption that all the fields involved are Gaussian, and is based on the theoretical ensemble average spectra, which are used to find the appropriate weights for the estimators to be unbiased and have the minimum variance. However, these properties of the QE lead to significant signal-to-noise penalties and biases in the high signal-to-noise regime. For example, the non-Gaussian nature of the optical depth  and galaxy density on small scales leads to the so-called $N^{(3/2)}$ bias which can exceed the Gaussian reconstruction noise in the high signal-to-noise regime~\cite{Giri:2020pkk}. For sufficiently low noise and high resolution CMB experiments, and in the absence of foregrounds, the dominant contribution to the observed temperature anisotropies on small angular scales will be the kSZ effect itself. In this limit, the QE is limited by sample variance, since it relies on searching for statistical anisotropies beyond those expected in the ensemble average. However, because one is able to accurately measure the detailed kSZ temperature anisotropies in this limit, it should be possible to do better by using the properties of our realization rather than the ensemble statistics (e.g. to use a form a sample variance cancellation). 

To overcome these two shortcomings of the QE, we introduce a new estimator for the radial velocity field based on a coarse-grained maximum likelihood condition. This maximum likelihood (MaxL) estimator does not rely on the assumption that the underlying fields are Gaussian (hence we do not expect the analogue of the $N^{(3/2)}$ bias), and relies on the realization of the underlying fields rather than their statistical properties (and hence should incorporate a form of sample variance cancellation). We therefore expect the MaxL estimator to have advantages over the QE in the high signal to noise regime, where the kSZ effect is signal dominated (e.g. over the primary CMB, instrumental noise, and foregrounds) in the CMB on small angular scales. Because it relies on the properties of a realization, forecasting with the MaxL estimator requires simulations, unlike the QE, whose statistical properties can be calculated analytically. We create properly correlated random Gaussian realizations of the underlying fields and make direct comparisons between the reconstruction of the QE as implemented previously~\cite{Cayuso:2021ljq} and the MaxL estimator. Because the realizations are purely Gaussian, we are unable to demonstrate the absence of a non-Gaussian bias for the MaxL estimator, and therefore it will be important in future work to compare the results of applying the MaxL estimator and QE to N-body simulations. 

A key ingredient of the MaxL estimator is an estimate of the optical depth field. In the absence of a direct measurement (e.g. by using a large number of Fast Radio Bursts~\cite{Madhavacheril:2019buy}), we must infer the optical depth from a tracer of large scale structure, such as a galaxy survey. This inference will be imperfect, leading to a multiplicative optical depth bias on the reconstructed velocities~\cite{Battaglia:2016xbi,Smith:2018bpn,Madhavacheril:2019buy,Hotinli:2021hih}. Further multiplicative biases will be introduced by redshift errors in the galaxy survey and various systematics in the galaxy and CMB surveys. These multiplicative biases on the QE are described in detail in Ref.~\cite{Cayuso:2021ljq}, and in this paper we describe their nature using a simplistic estimator for the optical depth field. Our implementation of the MaxL estimator can likely be improved upon through the use of more accurate methods to estimate the optical depth field, e.g. using machine learning techniques trained on accurate hydrodynamical simulations (see e.g.~\cite{Thiele:2020zpz}).  

The plan of the paper is as follows. In Sec.~\ref{sec:motivation} we provide some intuition and motivation for the MaxL estimator. In~Sec.~\ref{sec:kSZmodel} we review our model for the kSZ component of the temperature anisotropies, which is largely a review of Ref.~\cite{Cayuso:2021ljq}. In Sec.~\ref{sec:maxLestim}, we derive an implementation of the maximum likelihood estimator (given the optical depth field) for the radial velocity field in both harmonic space and map space. In Sec.~\ref{sec:forecasts} we compare the performance of the MaxL estimator and QE on a set of correlated Gaussian mock datasets. In Sec.~\ref{sec:tauestimate} we discuss inference of the optical depth field from a galaxy redshift survey and the associated optical depth bias on the reconstructed velocity; we conclude in Sec.~\ref{sec:conclusions}.  

\subsubsection*{ Some notation }

Throughout the paper, we refer to objects defined on a sphere in both in ``harmonic'' and ``map'' space. We label the multipole moments of fields in harmonic space by the pairs $(\ell , m),(\ell^\prime,m^\prime)\ldots$  and of the pixels in map space by the letters $i,j,\ldots$. For coarse-grained quantities, we use capital letters $(L,M),(L^\prime,M^\prime),\ldots$ or $I,J\ldots$. We denote operators in boldface, such that $\tau$ is a field but $\boldsymbol{\tau}$ is an operator.

\section{Intuition and motivation}\label{sec:motivation}

In this section we give a heuristic description of the QE that recovers the radial velocity field from a temperature and optical depth measurement, and motivate a new estimator that can avoid some of the QE reconstruction noise in the high-signal-to-noise regime of velocity reconstruction.

The kSZ contribution to the CMB temperature $\Theta$ is of the form $\tau v$, where $v$ is the radial velocity field and $\tau$ is the optical depth field, such that 
\begin{align}
\Theta =& \ \Theta^{\rm pCMB} +\Theta^{\rm kSZ}, \nonumber \\
\sim& \ \Theta^{\rm pCMB} + \tau v \label{heuristic_ksz},
\end{align}
where $\Theta^{\rm pCMB}$ describes all contributions to the temperature anisotropies that are not kSZ (this includes primary CMB contributions as well as other foreground contributions and instrumental noise). Here, we employ a toy model where kSZ is sourced by a two dimensional field; later we include line of sight effects. The goal is to use a measurement of $\Theta$ and $\tau$ to recover $v$. The QE recovers the velocity from this combination and their statistics:
\begin{equation}
\hat v^{\rm QE} \sim \frac{1}{{A}}\Theta \ \tau,
\end{equation}
where $A$ is a theoretically calculated filter that can be thought of as $\sim \left<\tau^2\right>$; it is calculated from the ensemble statistics of $\tau$. The QE differs from the true velocity by
\begin{align}
\hat v^{\rm QE}-v\sim& \frac{1}{\left<\tau^2\right>} \Theta \ \tau-v\nonumber\\
\sim & \ v \left(\frac{\tau^2}{\left<\tau^2\right>}-1\right)+\frac{\Theta^{\rm pCMB}}{\left<\tau^2\right>}.
\end{align}
In particular, there is a contribution to the residual sourced by the difference of the realization of $\tau^2$ from its theoretically expected variance. This contribution persists even when $\Theta^{\rm pCMB}$ is vanishingly small.

However, in the limit where the $\Theta^{\rm kSZ}$ contributions dominate over the non-kSZ contributions, we can avoid this contribution to the reconstruction noise by neglecting $\Theta^{\rm{pCMB}}$ and simply ``inverting'' Eq.~\eqref{heuristic_ksz}:
 \begin{equation}
\hat v\sim\frac{\Theta}{\tau}.
\end{equation}
In this case, the ``reconstruction noise'' only has contributions from the non-kSZ contributions to the temperature:
\begin{equation}
\hat v - v \sim\frac{\Theta^{\rm pCMB}}{\tau}.
\end{equation}
This allows us to make a lower-noise estimate of $v$ by utilizing angular scales where the kSZ temperature dominates over the other contributions to the CMB temperature, similar to what is done in the gradient-inversion approach to CMB lensing reconstruction~\cite{2019PhRvD.100b3547H}.

In the remainder of this Paper, we develop this idea more rigorously by deriving a new estimator from the maximization of a likelihood, and compare explicitly the performance of the ``MaxL'' estimator with that of the QE. Note that further complication arises from the fact that the kSZ temperature is not \textit{truly} sourced by the product of one velocity and one electron density field, as in Eq.~\eqref{heuristic_ksz}, but by the integral of this product along the line of sight. We will deal with this explicitly by ``redshift-binning'' the velocity and optical depth fields, with the end-goal being to reconstruct the redshift-binned velocity field.

\section{The kinetic Sunyaev Zel'dovich effect}\label{sec:kSZmodel}

The observed kSZ temperature anisotropies in the direction $\hat n$ are given by the line-of-sight integral over radial comoving distance $\chi$
\begin{equation}\label{eq:kszstart}
\Theta^{\rm kSZ} (\hat{n}) = \int_0^{\chi_{\rm max}} d\chi \ \dot{\tau} (\hat{n}, \chi) v(\hat{n}, \chi).
\end{equation}
Here, $v(\hat{n}, \chi)$ is the radial velocity field and the differential optical depth $\dot{\tau} (\hat{n}, \chi)$ is 
\begin{equation}
\dot{\tau} (\hat{n}, \chi) = - \sigma_T a(\chi) \bar{n}_e (\chi) \left( 1 + \delta_e (\hat{n}, \chi) \right),
\end{equation}
with $\sigma_T$ the Thompson cross section, $a(\chi)$ the scale factor, $\bar{n}_e (\chi)$ the cosmological mean electron density, and $\delta_e (\hat{n}, \chi) $ the electron overdensity field. There are contributions to the kSZ signal from during and after the epoch of reionization. We are primarily concerned with the late-time kSZ effect in this paper\footnote{See Ref.~\cite{Hotinli:2020csk} for a discussion of kSZ velocity reconstruction during the epoch of reionization.}, and so in Eq.~\eqref{eq:kszstart} we introduce a cutoff $\chi_{\rm max}$ on the line of sight integral at some time after reionization (which also justifies our neglect of the optical depth factor $e^{-\tau}$ in this equation). 

Following Ref.~\cite{Cayuso:2021ljq}, we decompose fields along the radial direction in the basis of Haar wavelets:
\begin{equation}
v(\hat{n}, \chi) = \sum_{s=0}^\infty v^s (\hat{n}) h^s (\chi),
\end{equation}
where $h^s$ are the Haar wavelet basis functions; these form a complete basis on the interval $0\leq \chi \leq \chi_{\rm max}$. The Haar expansion has the additional useful property that if we truncate at a fixed index $N$, the sum in the Haar basis is simply related to a sum over radial tophat bin-averages over bins of equal comoving size $\Delta \chi = \chi_{\rm max}/N$:
\begin{equation}
v(\hat{n}, \chi) = \sum_{\alpha=0}^{N-1} v^\alpha (\hat{n}) \Pi^\alpha (\chi) + \sum_{s=N}^\infty v^s (\hat{n}) h^s (\chi),
\end{equation}
where $\Pi^\alpha$ are radial top-hat bins. We will reserve Greek superscripts $\alpha, \beta, \gamma, \ldots$ to denote the tophat bin-averages. The bin averages $v^\alpha$ give a coarse-grained description of the field on the past light cone, and the remaining sum over Haar moments gives the residual fine-grained information. We can therefore write the line of sight integral in Eq.~\eqref{eq:kszstart} as the sum: 
\begin{equation}
\Theta^{\rm kSZ} (\hat{n}) =  \sum_{s=0}^{\infty} \tau^s (\hat{n}) v^s (\hat{n})  =  \sum_{\alpha=0}^{N-1} \tau^\alpha (\hat{n}) v^\alpha (\hat{n}) + \sum_{s=N}^{\infty} \tau^s (\hat{n}) v^s (\hat{n}),
\end{equation}
where we have defined $\tau^\alpha (\hat{n}) \equiv \Delta \chi \ \dot{\tau}^\alpha (\hat{n})$.
Below, we approximate the continuum limit by truncating at fixed $N$. For the contribution to kSZ from the redshift range $0< z< 5$ most of the kSZ autopower is captured for $N=512$ as shown in Ref.~\cite{Cayuso:2021ljq}. However, in practice we will have to truncate the sum at a somewhat lower value of $N$ for the application of the estimators derived below.

It will be useful to think of the optical depth as a set of $N$ operators that map the bin-averaged radial velocity field to the observed temperature anisotropies, in which case we can write the sky-basis-independent equation truncated at fixed $N$: 
\begin{equation}
\Theta^{\rm kSZ}  \simeq \sum_{\alpha=0}^{N-1} \boldsymbol{\tau}^\alpha \cdot v^\alpha.
\end{equation}
Here, just as any field defined on a sphere, $\Theta^{\rm kSZ}$ can be represented either in map space or  harmonic space. In map space, $\Theta_i$ has dimension equal to the number of pixels $N_{\rm pix}$ at some resolution (e.g. using the Healpix pixelization scheme~\cite{2005ApJ...622..759G}); $v_i^\alpha$ is a set of $N$ fields each with dimension equal to $N_{\rm{pix}}$; $\boldsymbol{\tau}_{ij}^\alpha$ is a set of $N$ operators each with dimension equal to $N_{\rm{pix}}\times N_{\rm{pix}}$. In harmonic space, $\Theta_{\ell m}$ has dimension equal to the number of modes required to describe the field at the appropriate resolution (which should in principle be equal to $N_{\rm{pix}}$, but in practice is somewhat larger); we denote this number as $N_{\rm harm}$. In Appendix~\ref{app:harmonic_map}, we present the form of $\boldsymbol{\tau}$ in each representation, as well as a mixed harmonic-map space representation.

\section{Maximum likelihood estimator}\label{sec:maxLestim}

The dominant blackbody components of the observed CMB temperature anisotropies are the lensed primary CMB and the late-time and reionization kSZ. Neglecting the reionization kSZ, the blackbody component of the temperature anisotropy is  given by:
\begin{equation}\label{eq:deltaT}
\Theta = \Theta^{\rm pCMB} + \sum_{s=0}^{\infty} {\tau}^s \cdot v^s.
\end{equation}
Here $\Theta^{\rm pCMB}$ are the primary CMB anisotropies in map or harmonic space, which we assume purely Gaussian with covariance $\boldsymbol{C}^{\mathrm{pCMB}}$. The likelihood function $\mathcal{L}$ for the primary CMB is then
\begin{equation}
\ln \mathcal{L} = (\Theta - \sum_{s=0}^{\infty} \boldsymbol{\tau}^s \cdot v^s )^\dagger (\boldsymbol{C}^{\mathrm{pCMB}})^{-1} (\Theta - \sum_{s=0}^{\infty} \boldsymbol{\tau}^s \cdot v^s) + \ln {\rm det} \ \boldsymbol{C}^{\mathrm{pCMB}}\label{likelihood}
\end{equation}
Extremizing $\mathcal{L}$ with respect to $(v^q)^\dagger$:
\begin{eqnarray}\label{eq:maxLcondition}
\frac{\delta}{\delta (v^q)^\dagger} \ln \mathcal{L}  = &(\boldsymbol{\tau}^q)^\dagger (\boldsymbol{C}^{\mathrm{pCMB}})^{-1} &(\Theta -  \sum_{s=0}^{\infty} \boldsymbol{\tau}^s \cdot v^s )= 0  \nonumber \\
\implies \ \ &(\boldsymbol{\tau}^q)^\dagger (\boldsymbol{C}^{\mathrm{pCMB}})^{-1} \Theta &= \sum_{s=0}^{\infty} \left[ (\boldsymbol{\tau}^q)^\dagger (\boldsymbol{C}^{\mathrm{pCMB}})^{-1} \boldsymbol{\tau}^s \right] \cdot v^s \nonumber \\
&(\boldsymbol{\tau}^\alpha)^\dagger (\boldsymbol{C}^{\mathrm{pCMB}})^{-1} \Theta &\simeq \sum_{\beta=0}^{N-1} \left[ (\boldsymbol{\tau}^\alpha)^\dagger (\boldsymbol{C}^{\mathrm{pCMB}})^{-1}\boldsymbol{ \tau}^\beta \right] \cdot v^\beta.
\end{eqnarray}
In the last line we truncate the radial sum at $N$ and work with the bin coefficients (denoted by the Greek indices). Note that $\Theta$ and $\boldsymbol{\tau}$ are fixed, and our goal is to find the set of $v^\alpha$ that maximize the likelihood. Organizing both $v^\beta$ and $(\boldsymbol{\tau}^\alpha)^\dagger (\boldsymbol{C}^{\mathrm{pCMB}})^{-1} \Theta$ into  $\left(N_{\rm pix}\times N\right)$-dimensional vectors containing both the bin and pixel information, and organizing $\left[ (\boldsymbol{\tau}^\alpha)^\dagger (\boldsymbol{C}^{\mathrm{pCMB}})^{-1} \boldsymbol{\tau}^\beta \right]$ into an $(N_{\rm pix} N) \times (N_{\rm pix} N)$-dimensional operator, Eq.~\eqref{eq:maxLcondition} is a linear equation for the velocity field:
\begin{equation}
\left[\boldsymbol{ \tau}^\dagger (\boldsymbol{C}^{\mathrm{pCMB}})^{-1} \Theta \right] =  \left[ \boldsymbol{\tau}^\dagger (\boldsymbol{C}^{\mathrm{pCMB}})^{-1} \boldsymbol{\tau} \right] \cdot v.\label{uncoarsegrained_maxl}
\end{equation}
This equation will have a solution when the operator $ \boldsymbol{\tau}^\dagger (\boldsymbol{C}^{\mathrm{pCMB}})^{-1} \boldsymbol{\tau} $ has a well-defined inverse. In general, this isn't the case, since there will be many highly oscillatory functions $v$ that nearly satisfy the maximum likelihood condition and pixels at which the operator (whose eigenbasis is in map space) is nearly zero (indicating small eigenvalues). However, we can first perform the matrix products in $ \boldsymbol{\tau}^\dagger (\boldsymbol{C}^{\mathrm{pCMB}})^{-1} \boldsymbol{\tau} $ at full resolution, and then coarse-grain this operator so that it is invertible. The maximum-likelihood condition is then
\begin{align}
&W\left(\left[\boldsymbol{ \tau}^\dagger (\boldsymbol{C}^{\mathrm{pCMB}})^{-1} \Theta \right]\right) =  W\left(\left[ \boldsymbol{\tau}^\dagger (\boldsymbol{C}^{\mathrm{pCMB}})^{-1} \boldsymbol{\tau} \right] \right)\cdot W\left(v\right)\nonumber\\
\implies W\left(v\right) = &\left[W\left(\left[ \boldsymbol{\tau}^\dagger (\boldsymbol{C}^{\mathrm{pCMB}})^{-1} \boldsymbol{\tau} \right] \right)\right]^{-1}W\left(\left[\boldsymbol{ \tau}^\dagger (\boldsymbol{C}^{\mathrm{pCMB}})^{-1} \Theta \right]\right)\label{cgmaxl}
\end{align}
where $W(\cdot)$ is the coarse-graining procedure; we demonstrate explicitly below that it is appropriate to separately coarse-grain $\left[ \boldsymbol{\tau}^\dagger (\boldsymbol{C}^{\mathrm{pCMB}})^{-1} \boldsymbol{\tau} \right]$ and $v$. The proposal is that with Eq.~\eqref{cgmaxl}, we can reconstruct the coarse-grained velocity $W(v)$. 
We now separately examine if there is a useful solution to the maximum likelihood condition in harmonic or map space for the velocity field. 
 
\subsection{Harmonic space estimator}
Before coarse graining, the maximum likelihood condition in harmonic space is
\begin{equation}\label{uncoarsegrained_maxl_harmonic}
\left[ \boldsymbol{\tau}^\dagger (\boldsymbol{C}^{\mathrm{pCMB}})^{-1} \Theta \right]_{\ell m}^\alpha = \sum_{\ell^\prime m^\prime; \ \beta} \left[ \boldsymbol{\tau}^\dagger (\boldsymbol{C}^{\mathrm{pCMB}})^{-1} \boldsymbol{\tau} \right]^{\alpha \beta}_{\ell m ; \ell^\prime m^\prime} v^\beta_{\ell^\prime m^\prime}.
\end{equation}
In Appendix~\ref{app:harmonic_map} we evaluate the components of the operators and vectors in Eq.~\eqref{uncoarsegrained_maxl_harmonic} explicitly, obtaining: 
\begin{equation}
\left[\boldsymbol{\tau}^\dagger (\boldsymbol{C}^{\mathrm{pCMB}})^{-1} \boldsymbol{\tau}\right]^{\alpha \beta}_{\ell m; \ell^\prime m^\prime} = \int d\Omega \ \tau^\alpha (\hat{n}) \bar{\tau}^\beta (\hat{n}) Y^*_{\ell m} (\hat{n}) Y_{\ell^\prime m^\prime} (\hat{n}) ,
\end{equation}
and
\begin{equation}
\left[ \boldsymbol{\tau}^\dagger (\boldsymbol{C}^{\mathrm{pCMB}})^{-1} \Theta \right]_{\ell m}^\beta = \int d\Omega \ \tau^\beta (\hat{n}) \bar{\Theta} (\hat{n}) Y^*_{\ell m} (\hat{n}) ,
\end{equation}
where the bar denotes inverse-variance filtering using $C_{\ell}^{\mathrm{pCMB}}$: 
\begin{equation}
\bar{A} (\hat{n}) \equiv \sum_{\ell m} \frac{A_{\ell m}}{C_{\ell}^{\mathrm{pCMB}}} Y_{\ell m} (\hat{n}).
\end{equation}

The coarse-graining operation required to make the operator $\left[ \boldsymbol{\tau}^\dagger (\boldsymbol{C}^{\mathrm{pCMB}})^{-1} \boldsymbol{\tau} \right]_{\ell m;\ell^\prime m^\prime}^{\alpha\beta} $ invertible is to cut off the sum at a maximum multipole $L_{\mathrm{max}}$, which will be the maximum multipole of the $v$ field that we recover; for more details, see Appendix~\ref{app:coarsegraining_harmonic}. After this coarse-graining, the maximum-likelihood solution for the velocity is:
\begin{equation}\label{eq:harmonicest}
\hat{v}_{LM}^\alpha = \sum_{L'M'}^{L_{\rm{max}}} \sum_{\beta=0}^{N-1} \left( [\boldsymbol{\tau}^\dagger (\boldsymbol{C}^{\mathrm{pCMB}})^{-1} \boldsymbol{\tau}]^{-1} \right)^{\alpha \beta}_{LM ; L'M'}  (\boldsymbol{\tau}^\dagger (\boldsymbol{C}^{\mathrm{pCMB}})^{-1} \Theta)_{L'M'}^\beta,
\end{equation}
where the $L_{\mathrm{max}}$ cut-off should be done after the calculation of $\boldsymbol{\tau}^\dagger (\boldsymbol{C}^{\mathrm{pCMB}})^{-1} \boldsymbol{\tau}$ but before its inversion.
The maximum values $N$ and $L_{\rm{max}}$ must be empirically determined from the numerical inversion of $\left[ \boldsymbol{\tau}^\dagger (\boldsymbol{C}^{\mathrm{pCMB}})^{-1} \boldsymbol{\tau} \right]_{\ell m;\ell^\prime m^\prime}^{\alpha\beta}$ for each realization. A crude approach to finding an approximate solution to the maximum likelihood condition Eq.~\eqref{uncoarsegrained_maxl_harmonic} is to substitute with the ensemble-average operator over realizations of the optical depth, which is explicitly invertible. This approach is presented in Appendix~\ref{sec:connectQE}, where we demonstrate that it yields an expression equivalent to the QE in the low signal-to-noise regime. We note that Eq.~\eqref{eq:harmonicest} is nearly identical to an estimator that was successfully used to reconstruct our peculiar velocity with respect the CMB rest frame using the thermal Sunyaev Zel'dovich effect~\cite{Planck:2020qil}.

We now compute the mean and variance of the MaxL estimator. The estimator Eq.~\eqref{eq:harmonicest} is based on a \textit{fixed realization} of the optical depth, so the statistics of the estimator should be computed over an ensemble of realizations of the primary CMB which we denote by $\langle \rangle_{\Theta}$. In this ensemble, the mean of the estimator is:
\begin{equation}
\langle \hat{v}_{LM}^\alpha \rangle_{\Theta} =  \sum_{L'M'}^{L'_{\rm{max}}} \sum_{\beta=0}^{N-1} \left( [\boldsymbol{\tau}^\dagger (\boldsymbol{C}^{\mathrm{pCMB}})^{-1} \boldsymbol{\tau}]^{-1} \right)^{\alpha \beta}_{LM ; L'M'}  \langle \left[\boldsymbol{\tau}^\dagger (C^{\mathrm{pCMB}})^{-1} \Theta \right]_{L'M'}^\beta \rangle_{\Theta} .
\end{equation}
We find that
\begin{equation}\label{harmonic_mean}
\langle \hat{v}_{LM}^\alpha \rangle_{\Theta} = v_{LM}^\alpha +  \beta_{LM}^\alpha,
\end{equation}
where $\beta$ is a ``coarse-graining'' bias 
\begin{equation}
\beta_{LM}^\alpha \equiv \sum_{L'M'}^{L_{\rm{max}}} \sum_{\gamma = 0}^{N-1} \sum_{c=N}^\infty \sum_{L''M''}^{L_{\rm{max}}}  \left( [\boldsymbol{\tau}^\dagger (\boldsymbol{C}^{\mathrm{pCMB}})^{-1} \boldsymbol{\tau}]^{-1} \right)^{\alpha \gamma}_{LM ; L'M'} [\boldsymbol{\tau}^\dagger (\boldsymbol{C}^{\mathrm{pCMB}})^{-1} \boldsymbol{\tau}]^{\gamma c}_{L'M' ; L''M''} v_{L''M''}^c.
\end{equation}
Details of the calculation that leads to~\eqref{harmonic_mean} are in Appendix~\ref{app:noise_covariance_harmonic}. We also calculate the variance of the estimator, and find that
\begin{equation}
\langle \hat{v}^\alpha_{LM} \hat{v}^\beta_{L'M'}  \rangle_{\Theta} =  (C^{vv})_{LM; L'M'}^{\alpha \beta} + N_{LM; L'M'}^{\alpha \beta} ,
\end{equation}
with
\begin{equation}
(C^{vv})_{LM; L'M'}^{\alpha \beta}  = (v_{L M}^\alpha + \beta_{L M}^\alpha) (v_{L' M'}^\beta +  \beta_{L' M'}^\beta) ,
\end{equation}
and
\begin{equation}
N_{LM; L'M'}^{\alpha \beta} =  \left( [\boldsymbol{\tau}^\dagger (\boldsymbol{C}^{\mathrm{pCMB}})^{-1} \boldsymbol{\tau}]^{-1} \right)^{\alpha \beta}_{LM ; L'M'} .
\end{equation}
Unlike the QE~\cite{Deutsch:2017ybc,Cayuso:2021ljq}, where one must assume that the underlying fields are Gaussian, it is not necessary to make any assumptions about the statistics of the optical depth or velocity fields. Therefore, we do not expect additional contributions to the estimator variance beyond what is presented here. This is in contrast to the QE, where extra contributions to the estimator variance arise from the non-Gaussian nature of the optical depth and velocities~\cite{Giri:2020pkk}. We view this as a significant advantage of the MaxL estimator, since these non-Gaussian contributions are difficult to compute and remove. 

Evaluating the harmonic space MaxL estimator in Eq.~\eqref{eq:harmonicest} requires costly, and potentially numerically inaccurate, forward and reverse spherical harmonic transforms. Motivated by this, in the next section we construct a map space estimator. Ultimately, the implementation of the map space estimator is simpler and more accurate, and will be the method used for the forecasts presented below. However, there may be contexts in which the harmonic space estimator is superior, as we comment on below. 
 
\subsection{Map space estimator}

We now derive an alternative coarse graining procedure and obtain a MaxL estimator defined in terms of fields in map space. In map space, (as demonstrated in Appendix~\ref{app:harmonic_map}) the maximum likelihood condition can be written explicitly as
\begin{equation}\label{uncoarsegrained_maxl_map_explicit}
 \tau^\alpha (\hat{n}) \bar{\Theta} (\hat{n}) =  \sum_{\beta=0}^{N-1} \tau^\alpha (\hat{n}) \bar{\tau}^\beta (\hat{n})  v^\beta (\hat{n}),
\end{equation}
where the bar denotes inverse-variance filtering by $C_{\ell}^{\mathrm{pCMB}}$:
\begin{equation}
\bar{A} (\hat{n}) \equiv \sum_{\ell m} \frac{A_{\ell m}}{C_{\ell}^{\mathrm{pCMB}}} Y_{\ell m} (\hat{n}).
\end{equation}
To coarse-grain, we define a set of smoothing kernels $W_I (\hat{n})$ with compact support in regions of the map labeled by $I=1,2, \ldots I_{\rm max}$ such that
\begin{equation}
\int d\Omega \ W_I (\hat{n}) W_J (\hat{n}) = \frac{\delta_{IJ}}{\Delta \theta^2}
\end{equation}
where we have assumed each kernel is non-zero over the same solid angle $\Delta \theta^2$ for all $I$. A simple choice for $W_I (\hat{n})$ satisfying this criterion, utilized below, is a simple average over equal-area pixels. Note that this is only one choice of definition for $W_I (\hat{n})$ among many possibilities. Multiplying both sides of Eq.~\eqref{uncoarsegrained_maxl_map_explicit} by $W_I (\hat{n})$ and integrating over the map, we obtain
\begin{equation}
\int d\Omega \ W_I(\hat{n}) \tau^\alpha (\hat{n}) \bar{\Theta} (\hat{n}) =  \sum_{\beta=0}^{N-1} \int d\Omega \ W_I(\hat{n}) \tau^\alpha (\hat{n}) \bar{\tau}^\beta (\hat{n})  v^\beta (\hat{n}).
\end{equation}
Expanding the velocity in terms of an average velocity $v^\beta_I$ over the region of the map labeled by $I$ and a residual $\Delta v^\beta_I (\hat n)$ in the same region, we have
\begin{equation}\label{eq:pixelML}
\int d\Omega \ W_I (\hat{n}) \tau^\alpha (\hat{n}) \bar{\Theta} (\hat{n}) =  \sum_{\beta=0}^{N-1} v^\beta_I \int d\Omega \ W_I (\hat{n}) \tau^\alpha (\hat{n}) \bar{\tau}^\beta (\hat{n}) +  \sum_{\beta=0}^{N-1} \int d\Omega \ W_I (\hat{n}) \tau^\alpha (\hat{n}) \bar{\tau}^\beta (\hat{n}) \Delta v^\beta_I (\hat n).
\end{equation}
Defining the operators
\begin{equation}
\Wop{I}{\alpha}{\beta} \equiv\int d\Omega \ W_I (\hat{n}) \tau^\alpha (\hat{n}) \bar{\tau}^\beta (\hat{n}),
\end{equation}
we have the following estimator for the average velocity in region $I$:
\begin{equation}
\hat{v}_I^\alpha = \sum_{\beta = 0}^{N-1} \Wop{I}{\alpha}{\beta}{}^{-1} \int d\Omega \ W_I (\hat{n}) \tau^\beta (\hat{n}) \bar{\Theta} (\hat{n}).
\end{equation}
In Appendix~\ref{sec:connectQE}, we demonstrate that the QE in the low signal-to-noise regime is recovered upon replacing $\Wop{I}{\alpha}{\beta} \rightarrow \langle \Wop{I}{\alpha}{\beta} \rangle_\tau$.

In our implementation of the map space estimator, we employ the Healpix equal-area pixelization scheme~\cite{2005ApJ...622..759G}. If we define the input fields at a native Healpix resolution with pixels indexed by $i = 1, 2, \ldots N_{\rm pix}^{\rm in}$ and the regions $I$ are defined by pixels at a degraded Healpix resolution $N_{\rm pix}^{\rm out}$, then we can write a discrete version of the MaxL estimator:
\begin{equation}
\hat{v}_I^\alpha = \sum_{\beta = 0}^{N-1} \left[  \sum_{i=1}^{ N_{\rm pix}^{\rm in}} W_{I i} \ \tau^\alpha_i \bar{\tau}^\beta_i \right]^{-1} \left[ \sum_{i=1}^{ N_{\rm pix}^{\rm in}} \ W_{I i} \ \tau^\beta_i  \bar{\Theta}_i \right].
\end{equation}
Note that $W_{Ii} $ as defined here is simply the Healpix \texttt{ud\_grade} function -- an operator that averages pixels at a fine resolution labeled by $i$ that fit within a coarse pixel labeled by $I$ (see Ref.~\cite{2005ApJ...622..759G}). 

Taking an ensemble average over realizations of the primary CMB, the estimator mean is
\begin{equation}
\langle \hat{v}_I^\alpha \rangle_{\Theta} = v_I^\alpha + \beta_I^\alpha\label{rmlmean}
\end{equation}
where the additive bias $\beta_I^\alpha$ arises from coarse graining:
\begin{eqnarray}
\beta_I^\alpha &=& \sum_{\beta; \gamma = 0}^{N-1} \Wop{I}{\alpha}{\beta}{}^{-1} \int d\Omega \ W_I (\hat{n}) \tau^\beta (\hat{n}) \bar{\tau}^\gamma (\hat{n}) \Delta v^\gamma_I (\hat{n}) \nonumber \\ 
&+&\sum_{\beta=0}^{N-1} \sum_{c=N}^\infty \Wop{I}{\alpha}{\beta}{}^{-1}\int d\Omega \ W_I (\hat{n}) \tau^\beta (\hat{n}) \bar{\tau}^c (\hat{n}) v^c (\hat{n}). \label{beta}
\end{eqnarray}
The first term is the bias from the pixel-averaging operation and the second term is the bias from coarse graining along the line of sight. Moving to the estimator variance:
\begin{eqnarray}
\langle \hat{v}_I^\alpha \hat{v}_J^\beta \rangle_{\Theta} &=& \left( v_I^\alpha + \beta_I^\alpha  \right) \left( v_J^\beta + \beta_J^\beta \right) + \frac{1}{\Delta\theta^2} \Wop{I}{\alpha}{\beta}{}^{-1}\delta_{IJ} \label{op_covariance}
\end{eqnarray}
Just as for the harmonic space estimator, the estimator variance includes an additive bias due to coarse graining as well as a reconstruction noise. Note that the reconstruction noise is local, and does not have any pixel-pixel covariance (although there is bin-bin covariance). A derivation of the estimator mean and variance can be found in Appendix~\ref{app:noise_covariance_map}. 

The map space estimator is local, only depending on the fields in the region supported by $W_I (\hat{n})$. This allows us to construct $I_{\rm max}$ independent $N \times N$ operators $\Wop{I}{\alpha}{\beta}$, which are more easily (and accurately) inverted than the single $(N_{\rm harm}^2 N) \times (N_{\rm harm}^2 N)$ operator needed for the harmonic space MaxL estimator. The local nature of the estimator also makes it a convenient choice for masked datasets. Indeed, one could generalize the discussion above to arbitrary complex pixelizations, suited to a particular procedure for masking, foreground removal, or anisotropic noise.

\section{Implementation and forecasts}\label{sec:forecasts}

In this section, we describe an implementation of the MaxL estimator and compare its performance to the existing QE introduced in Ref.~\cite{Deutsch:2017ybc}. Unlike the QE, whose performance can be forecast from theoretical spectra, the performance of the MaxL estimator depends on the realization. We therefore must generate ensembles of simulations to compare the performance of the MaxL estimator to the QE. In this paper, we perform reconstructions on ensembles of properly correlated Gaussian simulations generated using the algorithm of~\cite{Cayuso:2021ljq}; an analysis of N-body simulations is an important direction for future work. Specifically, we first calculate the theoretical power spectra of the lensed primary CMB $C_\ell^{pCMB}$; of the velocity field $C_\ell^{vv}$; and of the electron density field $C_\ell^{\tau \tau}$. We neglect the cross-power spectrum of electrons and velocity $C_\ell^{\tau v}$ since the estimators utilize the optical depth and velocities on very different scales. We use CAMB~\cite{Lewis:1999bs} to calculate quantities such as the primary CMB and linear matter power spectrum, with the non-linear regimes modelled with a halo model~\cite{Cayuso:2021ljq} (see e.g. Ref.~\cite{Cooray:2002dia} for a review of the halo model). The electrons are modelled with the density profiles of~\cite{2016JCAP...08..058B}. These power spectra are then used to create appropriately correlated Gaussian simulations of the binned velocity $v$ and $\tau$, as well as the primary CMB. To model the non-Gaussian kSZ map, we simply multiply and sum the simulated $v$ and $\tau$ maps: $\Theta^{\rm{kSZ}}=\sum v^\alpha \tau^\alpha$. We use 32 redshift bins between $z=0.2$ and $z=5$, and employ a binning scheme such that each bin has equal comoving width. Further details of this implementation are given in Sec. III of ~\cite{Cayuso:2021ljq}. We consider the idealized scenario where all non-blackbody foregrounds are absent and where we neglect the non-Gaussianities induced by lensing of the primary CMB. A discussion of these effects in the context of the QE can be found in Ref.~\cite{Cayuso:2021ljq}.

\begin{figure}[h!]
\includegraphics[width=0.5\textwidth]{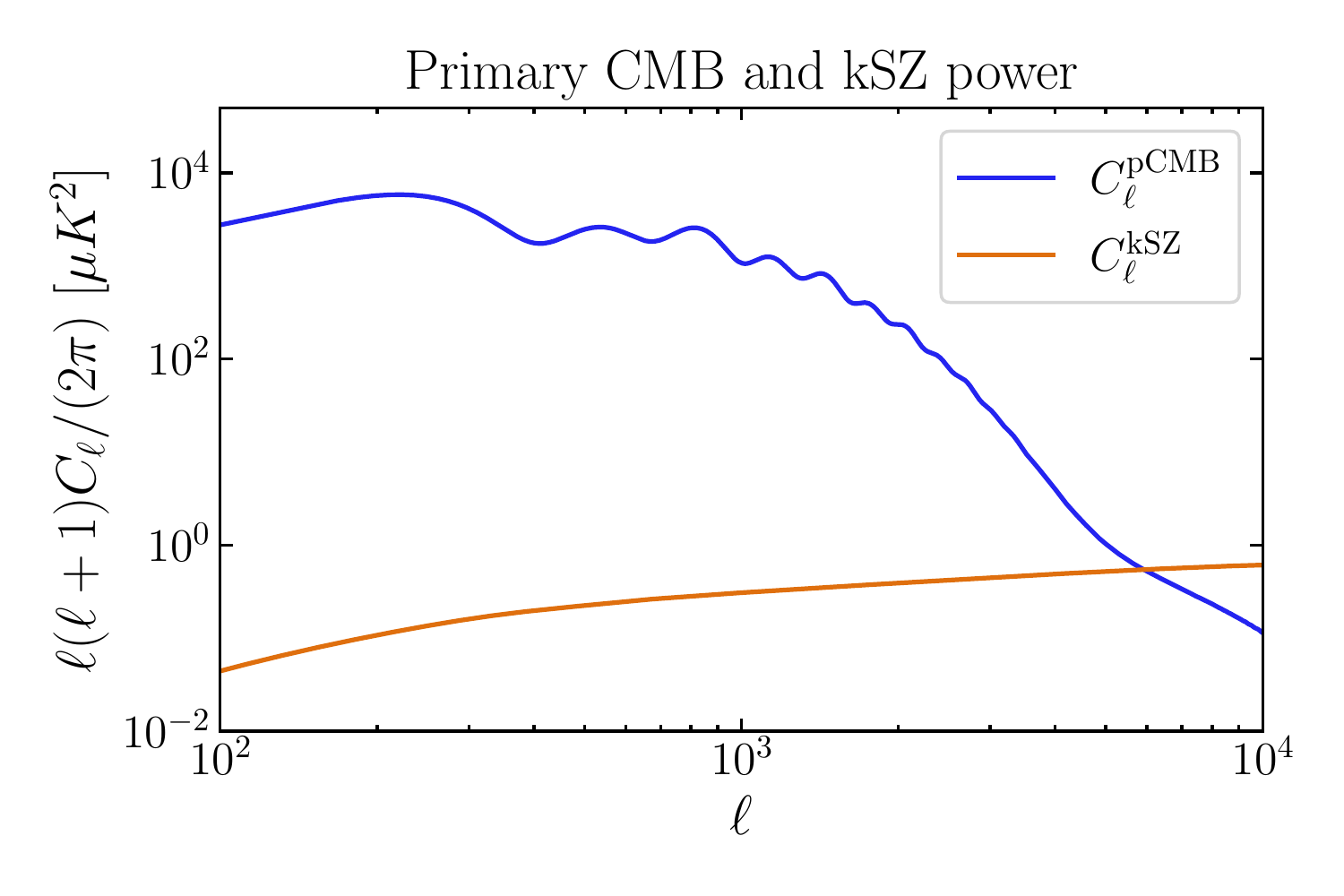}
\caption{The primary CMB power spectrum and the kSZ power spectrum using 32 radial bins of equal comoving size between $0.2 < z < 5$. At large-$\ell$, the kSZ power dominates over the CMB power. It is only once we access these angular scales that we expect the MaxL estimator to show improvements over the QE.}\label{fig:cmbkszpower}
\end{figure}

In Fig.~\ref{fig:cmbkszpower} we show the ensemble averaged power spectra of the (lensed) primary CMB and of the kSZ effect. Note that we only include power from the 32 redshift bins between $0.2 < z < 5$ used in our analysis, while Ref.~\cite{Cayuso:2021ljq} demonstrated that approximately $512$ radial bins are necessary to capture the majority of the kSZ power in this redshift range. We continue with this model in order to be self-consistent, as using a larger number of bins would be too computationally expensive. We do not expect this approximation to affect our conclusions. 

Throughout this section, we consider the case where we have access to a direct measurement of the optical depth. In practice, we will not have access to such a map: one must use a tracer of the electrons such as a galaxy survey. Connecting such a tracer to the optical depth necessitates further modelling, in particular through the cross power spectrum $C_\ell^{\tau g}$. Incorrect modelling of $C_\ell^{\tau g}$ leads to a bias on the reconstruction known as the optical depth bias; we comment on this further in Section~\ref{sec:tauestimate}. Additionally, we first consider the case where $C_\ell^{\tau  \tau}$ is known to arbitrarily high resolution, and don't consider realistic effects such as the shot noise of a galaxy survey or redshift error on a photometric survey. 

We have validated two pipelines, based the map and harmonic space MaxL estimators presented above.  We find that the map space estimator has a number of computational advantages. The harmonic space estimator involves several forward and reverse spherical harmonic transforms, which in the Healpix pixelization are not information-preserving. This loss of information leads to numerical inaccuracies in the reconstruction which can be overcome by using exact, but less efficient, pixelization schemes. In addition, the operator inversion required in the velocity estimator is costly when incorporating the full multipole and bin covariance; this places computational limitations on the degree of coarse graining that can be considered. The map space estimator overcomes both of these challenges, and so we focus on this implementation in what follows.

In our first illustration of the MaxL estimator, we completely remove the primary CMB and perform reconstruction on kSZ maps alone (we add a negligible amount of white noise to ensure that we are not dividing by zero when we filter). We use the equal-area Healpix pixelization~\cite{2005ApJ...622..759G}, where the resolution is set by the parameter $N_{\rm side}$, related to the number of pixels in the map by $N_{\rm pix} = 12 N_{\rm side}^2$. We apply both the MaxL estimator and the QE to the kSZ-only and optical depth maps for a single realization at a native resolution of $N_{\rm side}^{\rm in} = 2048$. Recalling the discussion in Sec.~\ref{sec:motivation}, we expect the QE to be limited by reconstruction noise, while the MaxL estimator is formally noiseless if the maximum likelihood condition can be inverted without coarse graining. Some degree of coarse graining is necessary for a well-defined solution to the maximum likelihood condition, and we therefore expect the MaxL estimator to be limited by the additive coarse graining bias as described in the previous section. 

\begin{figure}[h!]
\includegraphics[width=0.49\textwidth]{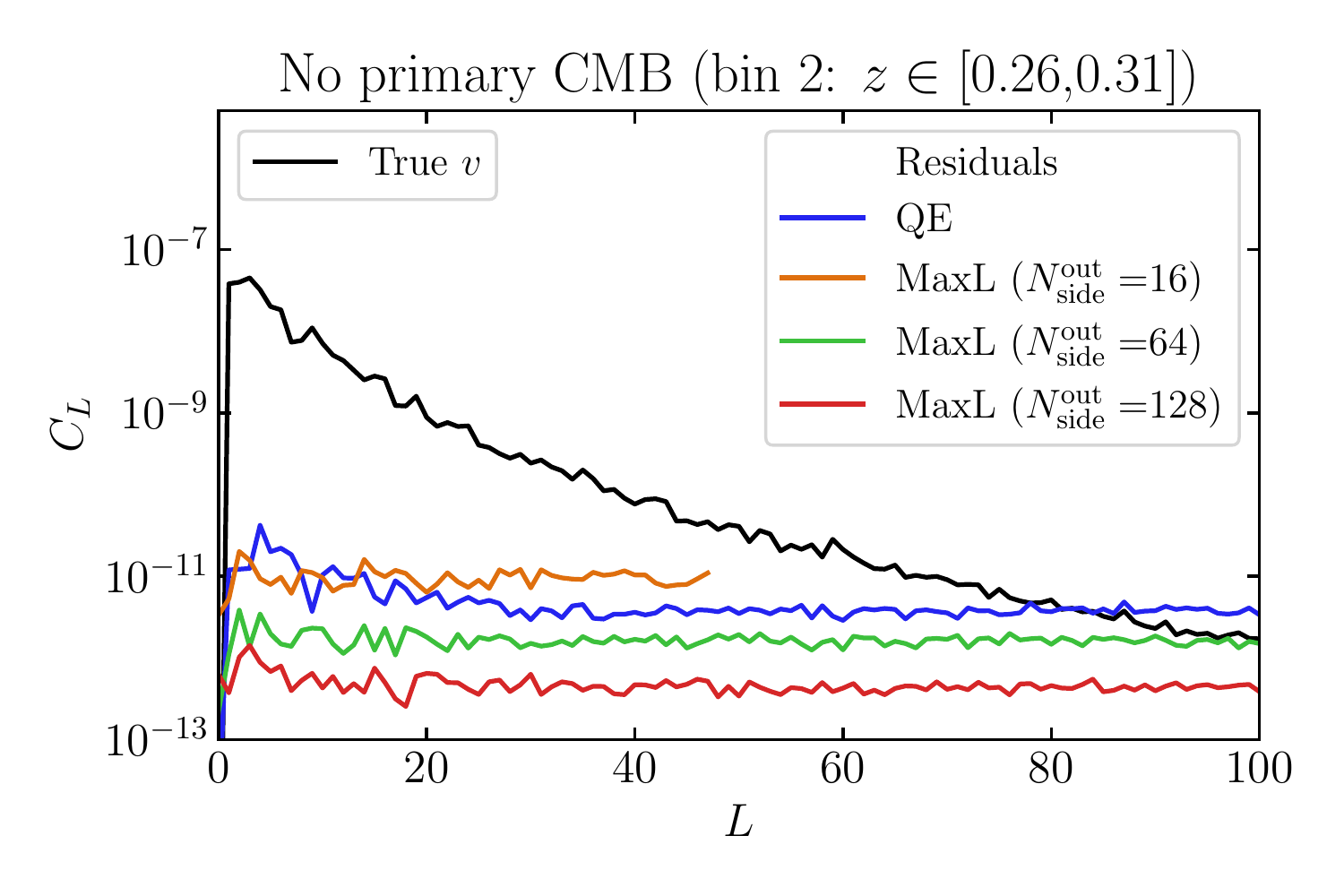}
\includegraphics[width=0.49\textwidth]{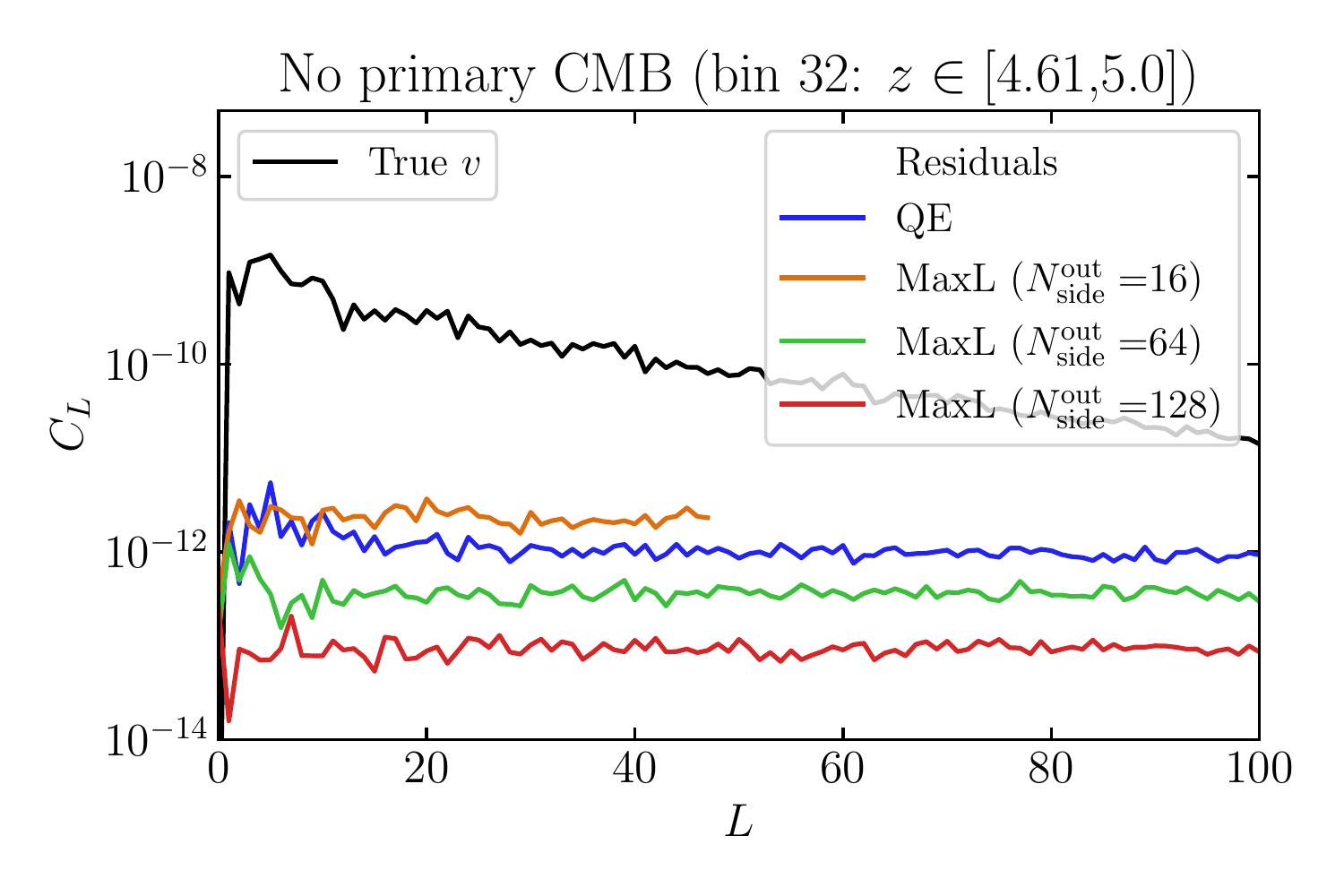}
\caption{We show the true velocity power spectrum as well as the power of the residuals $v-\hat v^{\rm{QE}}$ and $v-\hat v^{\rm{MaxL}}$. On the left is a low-$z$ bin and on the right is a high-$z$ bin. We show $\hat v^{\rm{MaxL}}$ for increasing resolution $N_{\rm side}^{\rm out}$. At low resolution, the ``coarse-graining'' bias $\beta$ is significant and $ \hat v^{\rm{MaxL}}$ performs worse than $\hat v^{\rm{QE}}$, but as we increase  $N_{\rm side}^{\rm out}$ this bias decreases and we get a better reconstruction than  $\hat v^{\rm{QE}}$. } \label{fig:perfect_reconstruction}
\end{figure}

The power spectrum of the true velocity and the power spectrum of the reconstruction residuals (defined by $\hat v-v$) are shown in Fig.~\ref{fig:perfect_reconstruction}. For the QE, we perform the reconstruction at a resolution of $N_{\rm side}^{\rm out} = 128$; for the MaxL estimator, we perform the reconstruction at a variety of resolutions $N_{\rm side}^{\rm out} = 16,64,128$. For low-resolution output maps ($N_{\rm side}^{\rm out} =16$) the MaxL estimator in fact performs worse than the QE due to a large coarse-graining bias $\beta$, but as we increase $N_{\rm side}^{\rm out}$ the MaxL estimator becomes superior as expected. We have explicitly checked that the residual agrees with $\beta$ as calculated directly from~\eqref{beta}.

\begin{figure}[h!]
\includegraphics[width=0.49\textwidth]{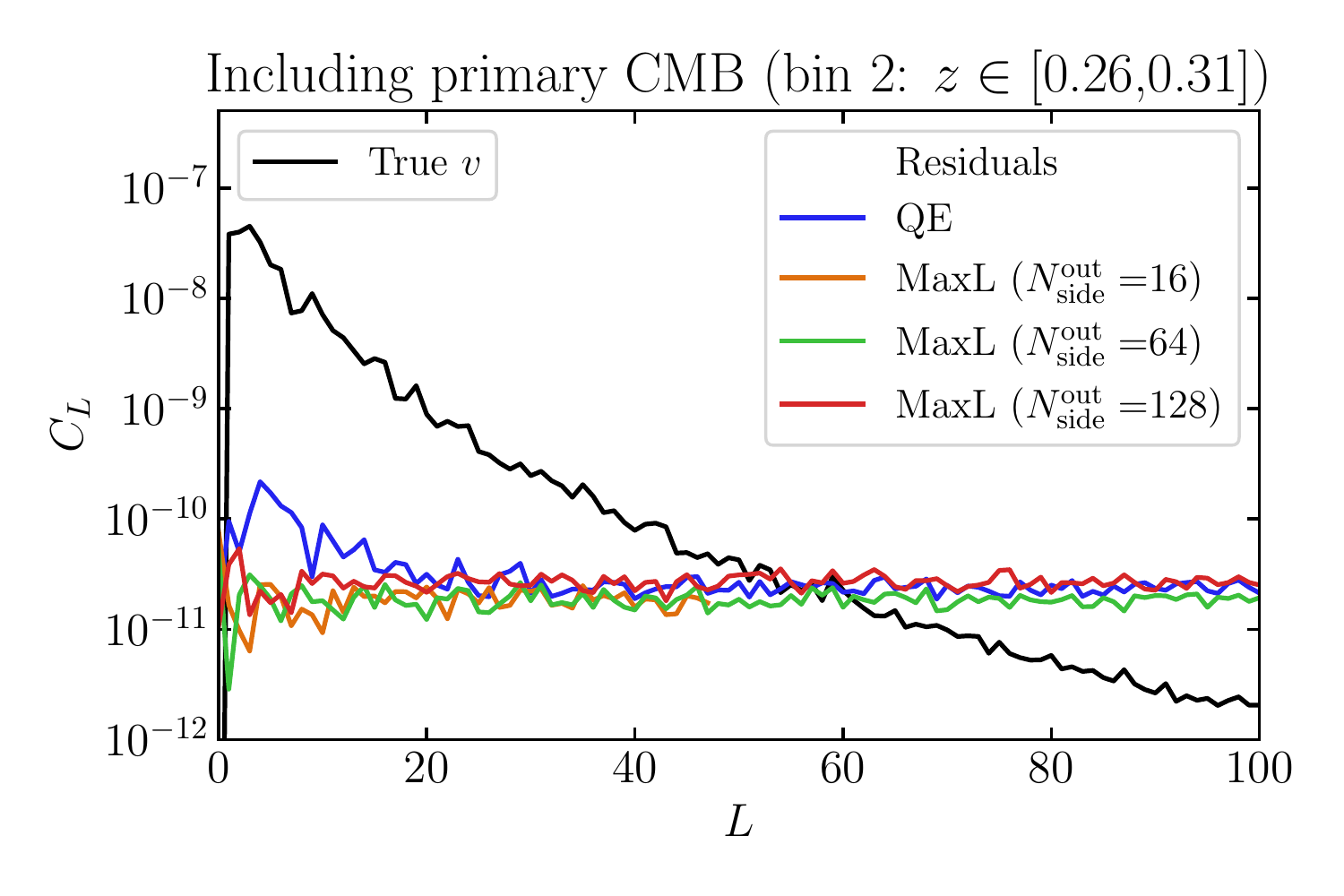}
\includegraphics[width=0.49\textwidth]{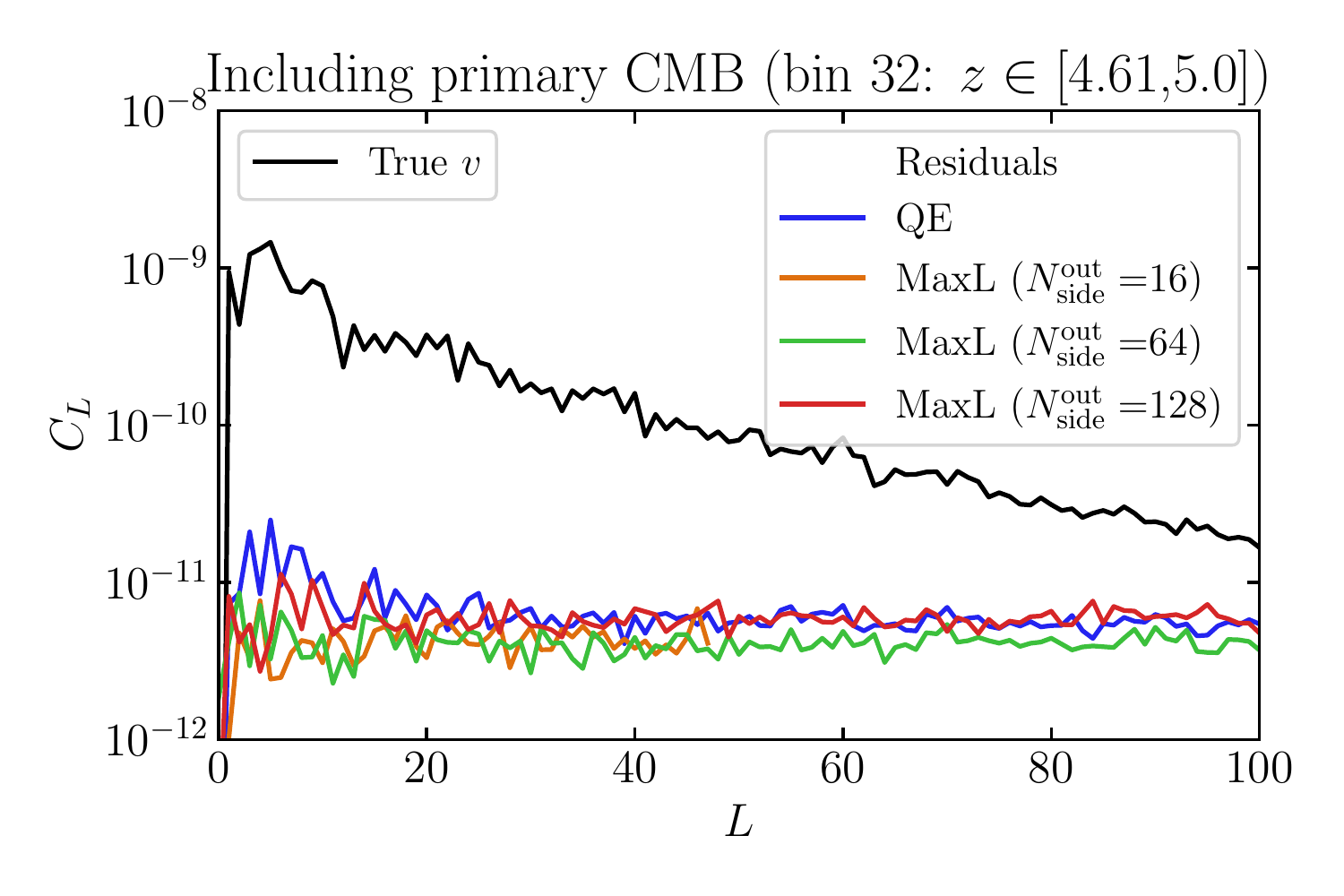}
\caption{We show the true velocity power spectrum as well as the power of the residuals $v-\hat v^{\rm QE}$ and $v-\hat v^{\rm MaxL}$. On the left is a low-$z$ bin and on the right is a high-$z$ bin. We show $\hat v^{\rm MaxL}$ for an increasing output resolution. Here we see less of an improvement over the QE than we did in Fig.~\ref{fig:perfect_reconstruction} where we did not include the primary CMB. The MaxL residual power is $~75 \%$ of the QE residual power on small angular scales, and significantly better on large angular scales. }\label{fig:CMB_reconstruction}
\end{figure}

\begin{figure}[h!]
\includegraphics[width=0.49\textwidth]{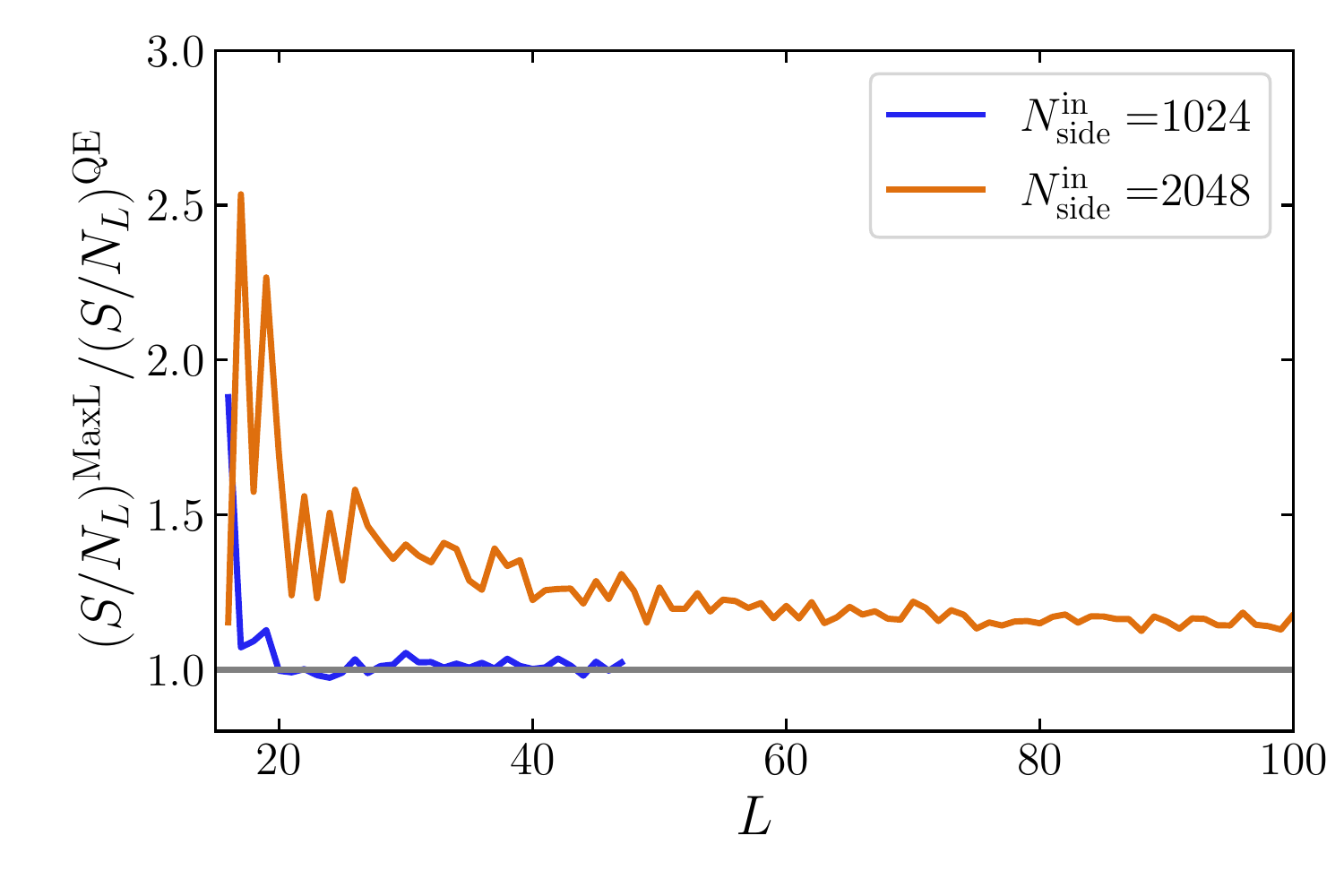}
\caption{The ratio of the total signal to noise per mode of the MaxL and quadratic estimators. We find that at $N_{\rm side}^{\rm in}$ of 1024, the estimators perform similarly; however, as we go to $N_{\rm{side}}^{\rm{in}}$ of 2048 and higher, the MaxL estimator has higher signal to noise than the QE. In both cases we have chosen a different $N_{\rm side}^{\rm out}$ at which the MaxL performs the best: if $N_{\rm side}^{\rm out}$ is too low, the coarse-graining bias is too large, and if it is to high, the operator $\mathcal W$ becomes un-invertible. These $N_{\rm side}^{\rm out}$ corresponded to $N_{\rm side}^{\rm out}=\{16,64\}$ for $N_{\rm side}^{\rm out}=\{1024,2048\}$ respectively. }\label{fig:s2n_comparison}
\end{figure}

When we include the primary CMB in the temperature maps, we do not expect the improvement of the MaxL estimator over the QE to be as dramatic. In this case, the MaxL estimator will have both reconstruction noise and an additive coarse-graining bias (see Eq.~\eqref{op_covariance}). We only expect the MaxL estimator to improve upon the QE when the resolution of the input maps is high enough to probe the regime where the amplitude of the kSZ anisotropies are comparable to those of the primary CMB. To test this expected resolution-dependence, we produce input maps at $N_{\rm side}^{\rm in}=1024, \ 2048$. In Fig.~\ref{fig:CMB_reconstruction} we compare the velocity reconstruction using the quadratic and MaxL estimators for $N_{\rm side}^{\rm out}  =16,64,128$ using maps with $N_{\rm side}^{\rm in} = 2048$. In Fig.~\ref{fig:s2n_comparison} we plot the ratio of the total signal to noise for the MaxL estimator and the QE; the signal to noise is defined as
\begin{equation}
S/N_L = \sqrt{\mathrm{Tr}\left[ C_L^{v^\alpha v^\beta} \left(N_L^{\hat{v}^\alpha\hat{v}^\beta}{}\right)^{-1}\right]}
\end{equation}
where $C_L^{v^\alpha v^\beta} $ is the (true) signal matrix and we use the power of the residual as a proxy for the noise $N_L^{\hat{v}^\alpha\hat{v}^\beta}$.

We see that the signal to noise for the MaxL estimator is comparable to that of the QE for $N_{\rm side}^{\rm in} = 1024$. Increasing to $N_{\rm side}^{\rm in} = 2048$, the MaxL signal to noise per mode is approximately 25\% higher than for the QE on small angular scales, and significantly higher on large angular scales. This demonstrates the improvement in performance expected once angular scales on which kSZ power is significant compared to the primary CMB are included. 

From these results we see that with perfect knowledge of the optical depth field and including only the primary CMB and kSZ temperature anisotropies, the MaxL estimator does yield a higher fidelity reconstruction than the QE for sufficiently high resolution data. In the next section, we take one step towards increasing realism and compare the performance of the MaxL and quadratic estimators when the optical depth is inferred from a galaxy survey.

\section{Estimating the optical depth}\label{sec:tauestimate}

In the absence of a direct measure of the differential optical depth, we must estimate it from a tracer in order to implement the maximum likelihood estimators introduced above. A simple estimator is:
\begin{equation}\label{eq:tauestimator}
\hat{\tau}_{\ell m}^\alpha = \sum_{\beta,\gamma=0}^{N-1} \left( C^{\tau g} \right)_{\ell}^{\alpha \beta} \left( \left[ C^{gg} \right]^{-1} \right)_{\ell}^{\beta \gamma}  g_{\ell m}^\gamma , 
\end{equation}
where $g_{\ell m}^\gamma$ is the redshift-binned galaxy density, $\left( C^{gg}  \right)_{\ell}^{\beta \gamma}$ is the galaxy power spectrum, and $\left( C^{\tau g} \right)_{\ell}^{\alpha \beta}$ is the cross-power with the optical depth. The ensemble mean of this estimator is zero while the variance is:
\begin{eqnarray}
\langle \hat{\tau}_{\ell m}^\alpha  (\hat{\tau}_{\ell m}^\beta)^\dagger \rangle 
&=& \sum_{\gamma,\delta=0}^{N-1} \left( C^{\tau g} \right)_{\ell}^{\alpha \gamma}\left( \left[ C^{gg} \right]^{-1} \right)_{\ell}^{\gamma \delta}  \left( C^{g \tau} \right)_{\ell}^{\delta \beta}  .
\end{eqnarray}
For correlated Gaussian fields Eq.~\eqref{eq:tauestimator} will yield an optical depth that has the expected statistics. However, it is likely that there are superior ways to faithfully estimate the optical depth from the galaxy density. We leave this question to future work. Note that the estimator variance reduces to $\left( C^{\tau \tau} \right)_{\ell}^{\alpha \beta}$ in the limit where the actual realization of the optical depth is related to the realization of the galaxy density by $\tau_{\ell m}^\alpha = \sum b_\ell^{\alpha \gamma} g_{\ell m}^\gamma$; that is, when the optical depth in a redshift bin can be obtained by isotropically filtering a linear combination of the tracers in different bins.  

An implementation of the MaxL estimator in this scenario involves two steps: first one estimates the optical depth field, then one uses in the MaxL estimator to obtain the reconstructed velocity. To compare the MaxL and quadratic estimators we must now include a galaxy field in our correlated Gaussian simulations. Following in the implementation in Ref.~\cite{Cayuso:2021ljq}, we generate galaxy auto- and cross-spectra for an LSST-like galaxy survey. We refer the reader to Ref.~\cite{Cayuso:2021ljq} for complete details, but in brief: the galaxy spectra are generated using a halo model incorporating a realistic halo occupation distribution (HOD), with number density of galaxies in the survey assumed to be
\begin{equation}
n(z) = \frac{n_g}{2 z_0} \left( \frac{z}{z_0} \right)^2 \exp \left( -\frac{z}{z_0} \right) ,
\end{equation}
with $z_0 = 0.3$ and $n_g = 40 \ {\rm arcmin}^{-2}$. The HOD and number density of objects determine the galaxy bias and galaxy shot noise in $\mathbf{C}_{\ell}^{gg}$; the halo model is used to determine a self-consistent $\mathbf{C}_{\ell}^{g\tau}$. Using these spectra as inputs, we generate properly correlated Gaussian maps of all fields at a Healpix resolution of $N_{\rm side}^{\rm in} = 2048$. Redshift errors associated with a photometric galaxy survey such as LSST lead to bin-bin correlations in the galaxy spectra on both large and small scales; we defer the inclusion of this effect to future work. 

\begin{figure}[h!]
\includegraphics[width=0.49\textwidth]{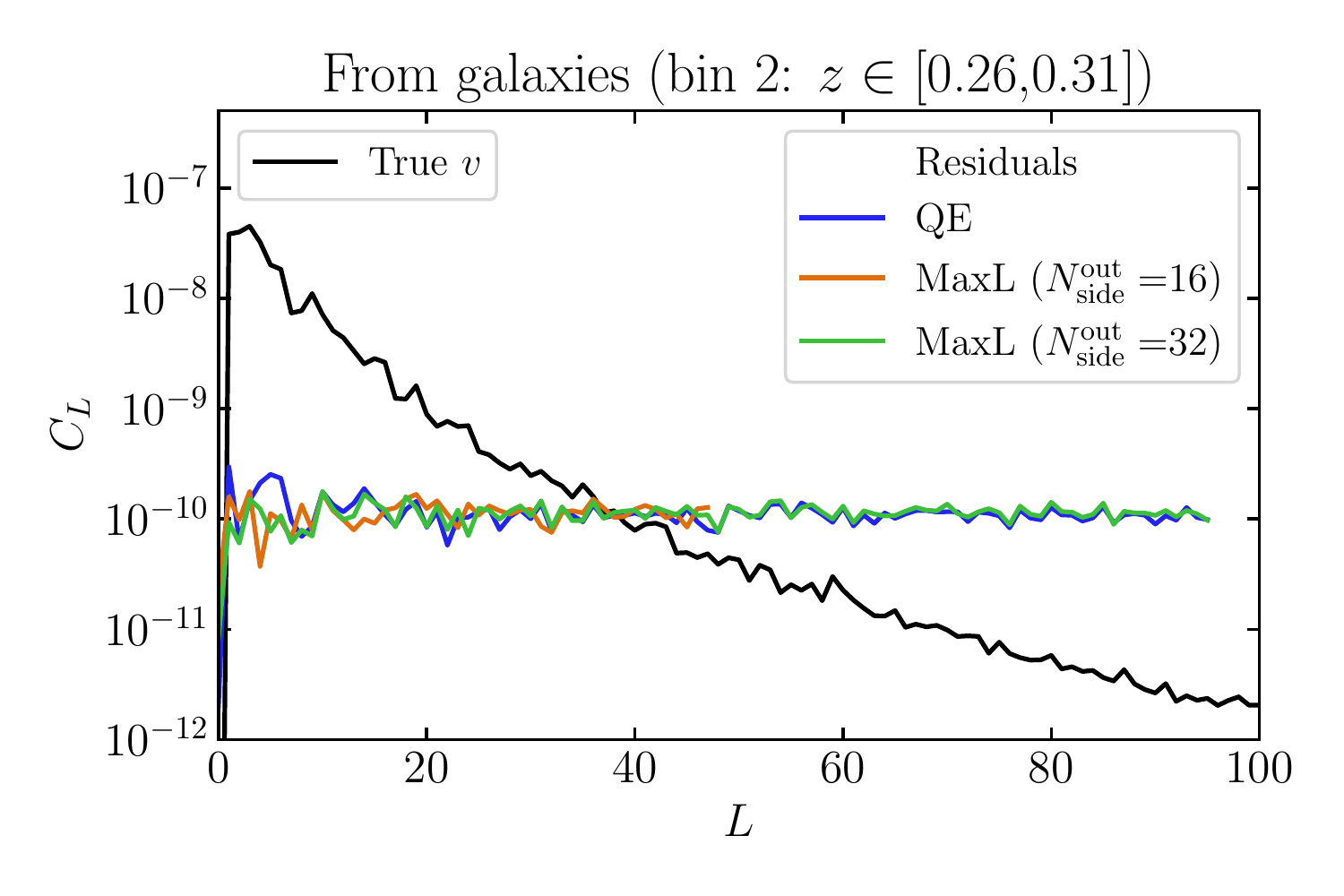}
\includegraphics[width=0.49\textwidth]{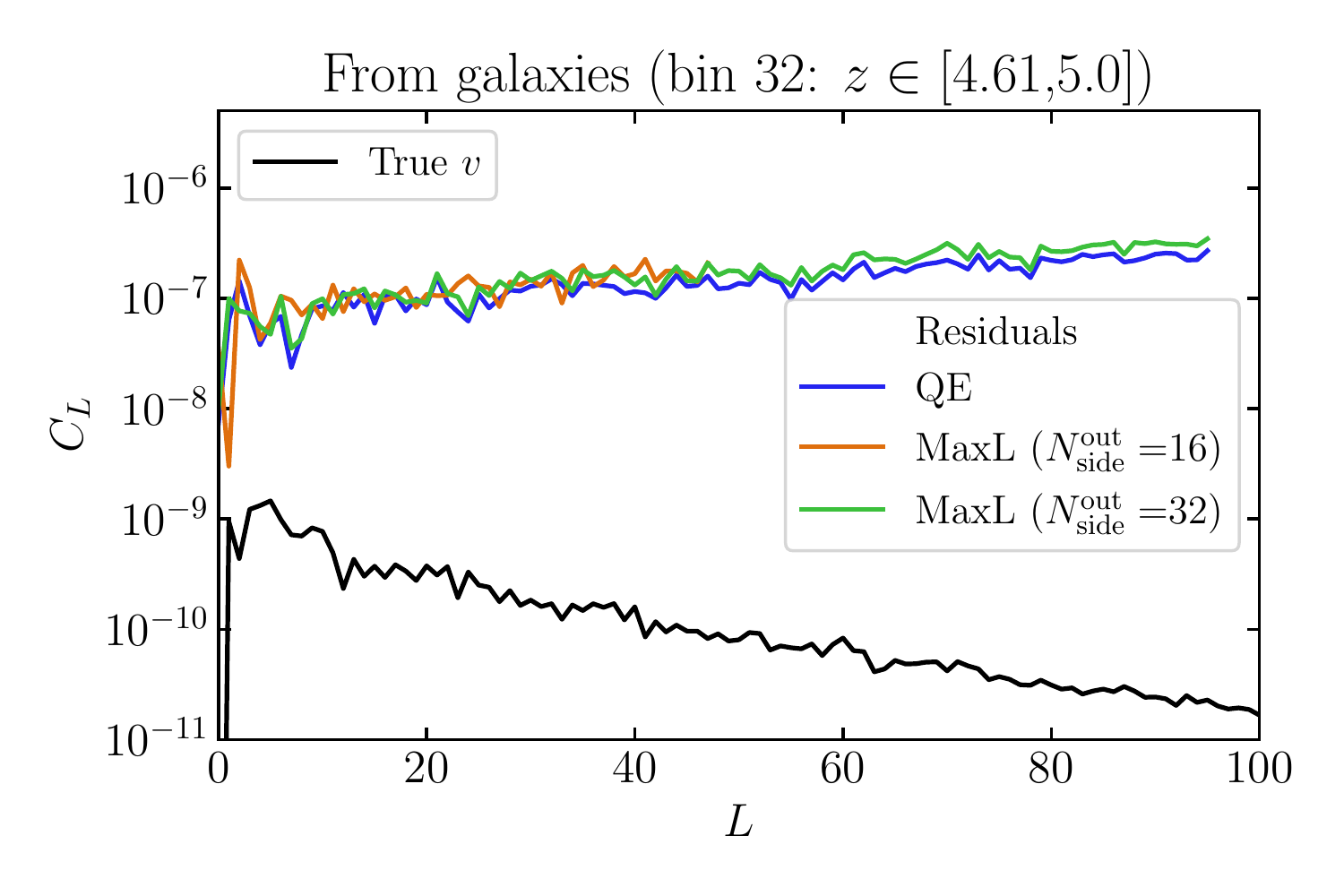}
\caption{In this Figure we show the true velocity power spectrum $C_L^{vv}$ and the power of the residuals $v-\hat v^{QE}$ and $v-\hat v^{MaxL}$, using galaxies to estimate the optical depth field. We do this for the MaxL by using the procedure defined in~\eqref{eq:tauestimator}. For the QE we use the simulated galaxy maps directly and modify the theoretical weights in the estimator to ensure an unbiased output.}\label{fig:galaxy_reconstruction}
\end{figure}

Using the Gaussian simulations described above, we implement the QE on the galaxy density and temperature (kSZ and primary CMB) maps using the theory spectra for $\mathbf{C}_{\ell}^{gg}$ and $\mathbf{C}_{\ell}^{g\tau}$ as inputs. We then use Eq.~\eqref{eq:tauestimator} to estimate the optical depth from the galaxy density and apply the MaxL estimator to the estimated optical depth and temperature map. The reconstruction residuals $\hat v-v$ for the MaxL and quadratic estimators are shown in Fig.~\ref{fig:galaxy_reconstruction}.  At low-redshfit (left panel), a high-fidelity reconstruction is possible, albeit at a lower signal-to-noise than what is possible using the optical depth itself as a tracer. At high redshift all galaxy density measurements are shot-noise dominated, as the galaxy density is very low. As such, it is not possible to reconstruct the high-$z$ velocity; this is evident on the right hand side of Fig.~\ref{fig:galaxy_reconstruction}. At both low- and high-redshift, the MaxL estimator performs at least as well as the QE, in this regime ($N^{\rm{in}}_{\rm{side}}=2048$). 
Given that an LSST-like experiment is shot-noise dominated on small angular scales, the additional information leveraged by the MaxL estimator from the optical depth on small scales is significantly degraded. Because of this, it may be difficult to identify an implementation of the MaxL estimator that would outperform the QE for near-term experiments. 

When a tracer is used in the MaxL estimator, we must confront the inevitable modelling uncertainty involved in estimating the optical depth field. In the estimator Eq.~\eqref{eq:tauestimator}, this modelling uncertainty arises through the theory spectra $\left(C^{\tau g} \right)_{\ell}^{\alpha \beta}$ and $\left( C^{gg} \right)_{\ell}^{\alpha \beta}$ that must be provided as inputs. As we demonstrate explicitly in Appendix~\ref{sec:opticaldepthbias}, the consequence is an overall multiplicative bias on the estimator mean, referred to in previous literature (e.g.~\cite{Battaglia:2016xbi,Smith:2018bpn,Madhavacheril:2019buy,Hotinli:2021hih}) as the optical depth bias. Neglecting the coarse-graining bias, for the map space estimator we have:
\begin{equation}
\langle \hat{v}_I^\alpha \rangle = \sum_{\gamma = 0}^{N-1} R^{\alpha \gamma}_I v_I^\gamma , 
\end{equation}
where the optical depth bias matrix is
\begin{equation}
R^{\alpha \gamma}_I \equiv \sum_{\beta = 0}^{N-1} \left[  \int d\Omega \ W_I (\hat{n}) \hat{\tau}^\beta (\hat{n}) \hat{\bar{\tau}}^\alpha (\hat{n}) \right]^{-1} \int d\Omega \ W_I (\hat{n}) \hat{\tau}^\beta (\hat{n}) \bar{\tau}^\gamma (\hat{n}) .
\end{equation}
A hat on $\tau$ implies that it has been reconstructed from a tracer, while the absence of a hat indicates the actual optical depth. In complete generality, the optical depth bias matrix contains $N^2 \times N_{\rm pix}$ free parameters. These free parameters are completely degenerate with the $N \times N_{\rm pix}$ independent degrees of freedom in the coarse grained velocity. Fortunately, as argued in Appendix~\ref{sec:opticaldepthbias}, we expect that the optical depth bias does not depend on the pixel index $I$. Roughly, this follows from the assumption that the optical depth field is statistically isotropic. Furthermore, we expect that the dominant contribution to the off-diagonal $\alpha \neq \gamma$ components of the optical depth matrix are expected to arise from redshift errors in the tracer used to reconstruct the optical depth. This implies that there are only $N$ free parameters in the optical depth bias matrix, so long as the redshift errors of the tracer can be well-characterized. This is consistent with the QE, where there are $N$ free parameters expected in the optical depth bias~\cite{Cayuso:2021ljq}. A complete investigation of the optical depth bias will be the focus of future work. 

\section{Conclusions}\label{sec:conclusions}

In this paper, we have introduced a new class of estimators for the radial velocity field (more generally, the remote dipole field) from the CMB temperature anisotropies induced by the kinetic Sunyaev Zel'dovich effect and a tracer of the optical depth field. These MaxL estimators are based on a coarse-grained maximum likelihood condition, and we have introduced two coarse-graining schemes in harmonic and map space that can be used to solve for the maximum likelihood radial velocity field. We derived the expected properties of the MaxL estimator, and compared the performance of the MaxL estimator to the existing quadratic estimator using the correlated Gaussian simulation framework of Ref.~\cite{Cayuso:2021ljq}.

The MaxL estimator introduced here has several advantages over the existing QE for velocity reconstruction. As we have demonstrated, the MaxL estimator can in principle yield a higher fidelity reconstruction than the QE in the high-resolution, low-noise frontier of future CMB experiments. The MaxL estimator is not derived based on the assumption that the optical depth field is Gaussian, and it therefore does not receive contributions from the non-Gaussian bias terms present in the QE~\cite{Giri:2020pkk}. The map space version of the MaxL estimator is numerically stable and simple to implement. It is absolutely local, making it a useful and versatile tool to apply to datasets containing masks and spatially varying foregrounds. The modular nature of the MaxL estimator makes it amenable to improvements in techniques used to estimate the optical depth field or improved coarse-graining schemes. Furthermore, applying both the quadratic and MaxL estimators to future datasets will be a useful diagnostic of various biases and systematics in each analysis.  

There are a variety of directions to take in future work. An important first step is a comparison of the quadratic and MaxL estimators using N-body data, where some of the advantages of the MaxL estimator described above can be confirmed in a more realistic setting. Another line of inquiry is to determine better methods for estimating the optical depth field from a galaxy redshift survey, or other tracers of large scale structure such as the cosmic infrared background~\cite{McCarthy:2019xwk}. For example, machine learning techniques trained on simulations could capture  non-Gaussian information to better model the tracer-optical depth correlations; such a framework could incorporate multiple tracers into an overall estimate of the optical depth field over a wide range of redshifts. It will also be important to develop an analysis framework incorporating a mask and spatially varying foreground mitigation techniques as exists for the QE~\cite{Cayuso:2021ljq}. Finally, one can construct an analogous set of MaxL estimators based on other CMB secondaries; for example, transverse velocity reconstruction from the moving lens effect~\cite{Hotinli:2018yyc} or remote quadrupole reconstruction from the polarized Sunyaev Zel'dovich effect~\cite{Deutsch_2018,Deutsch:2017ybc}. Improved techniques to produce unbiased high-fidelity reconstruction of cosmological fields will yield significant scientific dividends for future experiments, maximizing the ability of these datasets to constrain fundamental properties of our Universe.

\acknowledgements
We thank R. Bloch, C. Hill, S. Hotinli, M. Madhavacheril, J. Mertens, M. M\"{u}nchmeyer, and B. Sherwin for useful conversations and input at various stages of this project. MCJ is supported by the National Science and Engineering Research Council through a Discovery grant. This research was supported in part by Perimeter Institute for Theoretical Physics. Research at Perimeter Institute is supported by the Government of Canada through the Department of Innovation, Science and Economic Development Canada and by the Province of Ontario through the Ministry of Research, Innovation and Science. Some of the results in this paper have been derived using the HEALPix package~\cite{2005ApJ...622..759G}.

\bibliography{references}

\appendix

\section{Harmonic and map space representations}\label{app:harmonic_map}

In this paper, we work both in harmonic space and in map space (also referred to as ``pixel'' or ``real'' space). In each case, we are considering fields in a different basis: in map space, we describe fields by associating a value to each pixel on a sphere, while in harmonic space we describe fields by expanding in the basis of spherical harmonics. In this Appendix, we briefly recall how to transform from one basis to another, considering both the fields themselves, and the operators on these fields.

\subsection{Fields}
A 2-dimensional field $A(\hat n)$ defined on a sphere can be represented in the map or harmonic basis:
\begin{equation}
a(\hat n) = \sum_i a_i P_i(\hat {n}) = \sum _{\ell m}a_{\ell m} Y_{\ell m}(\hat{ n})
\end{equation}
where $a_i$ is the value of the field at each pixel $P_i(\hat n)$ and $a_{\ell m}$ are the spherical harmonic coefficients of the field decomposed onto the spherical harmonic basis $Y_{\ell m}( \hat n)$. 
In the map basis, each pixel can be denoted $i$ and the value $a_i$ is given by
\begin{equation}
a_i = \int d\Omega \ a(\hat n) P_i(\hat {n})
\end{equation}
where $P_i$ is an element of the discrete pixelization basis. In the harmonic basis, fields can be represented as a sum over the spherical harmonic basis $Y_{\ell m}$ with the $a_{\ell m}$'s given by
\begin{equation}
a_{\ell m}=\int d\Omega \ a(\hat n) Y_{\ell m}^*(\hat n).
\end{equation}

In particular, we can go between map-space and harmonic-space according to
\begin{align}
a_{\ell m}=&\int d\Omega\left( \sum_i a_i P_i (\hat {n}) \right)Y_{\ell m}^*(\hat n)=\sum_i a_i  Y_{\ell m}^*(\hat n_i);\\
a_i =& \int d\Omega \sum _{\ell m}a_{\ell m} Y_{\ell m}(\hat{ n}) P_i(\hat {n})=\sum_{\ell m} a_{\ell m}  Y_{\ell m}(\hat n_i),
\end{align}
where $ Y_{\ell m}(\hat n_i) \equiv \int d\Omega P_i (\hat {n}) Y_{\ell m}(\hat n)$, e.g. the spherical harmonic weighted over the pixel centered on $\hat n_i$.

\subsection{Operators}

We can decompose operators $\mathbf{O}$ on either basis. Thus the expression
\begin{equation}
\boldsymbol{O} b = c
\end{equation}
can be written as 
\begin{equation}
\sum_j\boldsymbol{O}_{ij} b_j = c_i,
\end{equation}
or
\begin{equation}
\sum_{LM}\boldsymbol{O}_{\ell m; L M} b_{L M} = c_{\ell m}.
\end{equation}
We can also consider a mixed basis
\begin{equation}
\sum_{LM} \boldsymbol{O}_{i; LM}b_{LM} = c_i.
\end{equation}
We also have the relation
\begin{align}
\boldsymbol{O}_{\ell m; L M} b_{L M} = &\sum_i\boldsymbol{O}_{ij} b_j  Y_{\ell m}^*(\hat n_i)\\
=&\sum_i\boldsymbol{O}_{ij} \sum_{\ell^\prime m^\prime }b_{\ell^\prime m^\prime} Y_{\ell^\prime m^\prime}(\hat n_j) Y_{\ell m}^*(\hat n_i).\label{operator_map2lm}
\end{align}

\subsection{The maximum likelihood condition in map and harmonic space}\label{app:harmonicmap_maxl}

The maximum likelihood condition is 
\begin{equation}\label{eq:appmaxLcond}
\left[\boldsymbol{ \tau}^\dagger (\boldsymbol{C}^{\mathrm{pCMB}})^{-1} \Theta \right] =  \left[ \boldsymbol{\tau}^\dagger (\boldsymbol{C}^{\mathrm{pCMB}})^{-1} \boldsymbol{\tau} \right] \cdot v.
\end{equation}
To write this in harmonic or map space, we need to form both the operator $  \boldsymbol{\tau}^\dagger (\boldsymbol{C}^{\mathrm{pCMB}})^{-1} \boldsymbol{\tau} $ and the vector $\boldsymbol{ \tau}^\dagger (\boldsymbol{C}^{\mathrm{pCMB}})^{-1} \Theta  $ in map and harmonic space. Let us first consider the operator  $  \boldsymbol{\tau}^\dagger (\boldsymbol{C}^{\mathrm{pCMB}})^{-1} \boldsymbol{\tau} $.

First, it will be useful to write the elements of $\boldsymbol{\tau}$ and $\boldsymbol{C}^{\mathrm{\mathrm{pCMB}}}$. In map space, $\boldsymbol{\tau}$ is diagonal:
\begin{equation}
\boldsymbol{\tau}^\alpha_{ij} = \tau^\alpha_i \delta_{ij}
\end{equation}
where $\tau^\alpha_i$ are the components of the $\tau^\alpha$ field in the position basis and $\delta_{ij}$ is the Kronecker delta, the identity on map space. Using the basis-transformation Eq.~\eqref{operator_map2lm} for a diagonal operator on map space gives
\begin{equation}
\boldsymbol{\tau}^\alpha_{\ell m; L M}  = \sum_{\ell^\prime m^\prime}\tau^\alpha_{\ell^\prime m^\prime}\int d\Omega \ Y_{\ell^\prime m^\prime}(\hat n) Y_{LM}(\hat n) Y_{\ell m}^*(\hat n).
\end{equation}
We can also write a mixed-basis expression
\begin{equation}
\boldsymbol{\tau}^\alpha_{i; L M} = \tau^\alpha_i Y_{LM}(\hat n_i).
\end{equation}

It will be most helpful to consider $\boldsymbol{C}^{\mathrm{\mathrm{pCMB}}}$ in harmonic space, where it is diagonal:
\begin{equation}
\boldsymbol{C}^{\mathrm{\mathrm{pCMB}}}_{\ell m ; \ell^\prime m^\prime} = C_\ell^{\mathrm{\mathrm{pCMB}}} \delta_{\ell \ell^\prime }\delta_{m m^\prime} ,
\end{equation}
where $C_\ell^{\mathrm{\mathrm{pCMB}}} $ is the power spectrum of the primary CMB. Then we can evaluate
\begin{eqnarray}
\left[ (\boldsymbol{C}^{\mathrm{pCMB}})^{-1} \boldsymbol{\tau}\right]^\alpha_{\ell^{\prime \prime} m^{\prime \prime}LM}&=&\sum_{\ell m}(\boldsymbol{C}^{\mathrm{pCMB}})^{-1}_{\ell^{\prime \prime} m^{\prime \prime} \ell m} \boldsymbol{\tau}^\alpha_{\ell m ;LM}\nonumber\\
 &=&  \sum_{\ell m; \ \ell' m'} \left[ \int d \Omega \ Y_{\ell' m'} Y_{LM} Y^*_{\ell m} \right] \frac{\tau^\alpha_{\ell' m'}}{C_\ell^{\mathrm{pCMB}}} \delta_{\ell \ell''} \delta_{m m''} \\
&=& \sum_{\ell' m'} (-1)^{m''} \sqrt{\frac{(2\ell'+1) (2L+1) (2 \ell''+1)}{4\pi}}
\begin{pmatrix}
L & \ell' & \ell'' \\
0 & 0 & 0
\end{pmatrix}
\begin{pmatrix}
L & \ell' & \ell'' \\
M & m' & -m''
\end{pmatrix}
          \frac{\tau^\alpha_{\ell' m'}}{C_{\ell''}^{\mathrm{pCMB}}} \nonumber
\end{eqnarray}
where $\begin{pmatrix}
L & \ell' & \ell'' \\
M & m' & -m''
\end{pmatrix}$ is the Wigner $3J$-symbol. 
Without making further assumptions, this is as far as we can go. However, we will mainly be interested in the regime where $L \ll \ell', \ \ell''$. The triangle inequality restricts $\ell$ and $\ell'$ to the range $| \ell''-\ell' | \leq L \leq \ell''+\ell'$. So, as long as $C_{\ell''}^{\mathrm{pCMB}}$ is not a strongly varying function, we can approximate:
\begin{equation}
C_{\ell''}^{\mathrm{pCMB}} \simeq C_{\ell'' \pm L}^{\mathrm{pCMB}} \simeq C_{\ell'}^{\mathrm{pCMB}}.
\end{equation}
Under this approximation, we have:
\begin{eqnarray}
\sum_{\ell m}\left[(\boldsymbol{C}^{\mathrm{pCMB}})^{-1}\right]_{\ell'' m''; \ell m} \boldsymbol{\tau}^\alpha_{\ell m; L M} &=& \int d \Omega \ \left[\sum_{\ell' m'} \frac{\tau^\alpha_{\ell' m'}}{C_{\ell'}^{\mathrm{pCMB}}} Y_{\ell' m'}\right] Y_{LM} Y^*_{\ell'' m''} ,
\end{eqnarray}
which has a nice mixed pixel-space harmonic-space representation:
\begin{eqnarray}
\left[(\boldsymbol{C}^{\mathrm{pCMB}})^{-1}\right]_{i j} \boldsymbol{\tau}^\alpha_{j; L M} 
&=& \bar{\tau}^\alpha_i Y_{LM} (\hat{n}_i)
\end{eqnarray}
where $\bar{\tau}^\alpha_i$ is the inverse-variance filtered field
\begin{equation}
\bar{\tau}^\alpha_i \equiv \sum_{\ell' m'} \frac{\tau^\alpha_{\ell' m'}}{C_{\ell'}^{\mathrm{pCMB}}} Y_{\ell' m'} (\hat{n}_i).
\end{equation}
With these results, we can now form the operator $\left[ \boldsymbol{\tau}^\dagger (\boldsymbol{C}^{\mathrm{pCMB}})^{-1}\boldsymbol{ \tau} \right]$. It is most convenient to do this using the mixed representation:
\begin{equation}
\left[\boldsymbol{\tau}^\dagger (\boldsymbol{C}^{\mathrm{pCMB}})^{-1} \boldsymbol{\tau}\right]^{\alpha \beta}_{L'M' ; LM} = \int d\Omega \ \tau^\alpha (\hat{n}) \bar{\tau}^\beta (\hat{n}) Y_{L'M'} (\hat{n})^* Y_{LM} (\hat{n}).
\end{equation}
Using an analogous set of arguments, we have:
\begin{equation}
\left[ \boldsymbol{\tau}^\dagger (\boldsymbol{C}^{\mathrm{pCMB}})^{-1} \Theta \right]_{L'M'}^\beta = \int d\Omega \ \tau^\beta (\hat{n}) \bar{\Theta} (\hat{n}) Y_{L'M'}^*  (\hat{n})
\end{equation}
where, again, the bar denotes inverse-variance filtering:
\begin{equation}
\bar{\Theta}_i \equiv \sum_{\ell' m'} \frac{\Theta_{\ell' m'}}{C_{\ell'}^{\mathrm{pCMB}}} Y_{\ell' m'} (\hat{n}_i).
\end{equation}

The maximum likelihood condition is
\begin{equation}
\left[ \boldsymbol{\tau}^\dagger (\boldsymbol{C}^{\mathrm{pCMB}})^{-1} \Theta \right]_{L'M'}^\alpha = \sum_{LM; \ \beta} \left[ \boldsymbol{\tau}^\dagger (\boldsymbol{C}^{\mathrm{pCMB}})^{-1} \boldsymbol{\tau} \right]^{\alpha \beta}_{L'M' ; LM} v^\beta_{LM}.
\end{equation}
We can also rotate into map space by first writing
\begin{equation}
\left[ \boldsymbol{\tau}^\dagger (\boldsymbol{C}^{\mathrm{pCMB}})^{-1}\boldsymbol{ \tau} \right]^{\alpha \beta}_{i ; LM} =  \tau^\alpha_i \bar{\tau}^\beta_i \ Y_{LM} (\hat{n}_i)   
\end{equation}
and
\begin{equation}
\left[\boldsymbol{ \tau}^\dagger (\boldsymbol{C}^{\mathrm{pCMB}})^{-1} \Theta \right]_{i}^\alpha = \tau^\alpha_i \bar{\Theta}_i.
\end{equation}
The maximum likelihood condition is then:
\begin{equation}\label{eq:map_ML}
 \tau^\alpha_i \bar{\Theta}_i = \sum_{LM; \ \beta} \tau^\alpha_i \bar{\tau}^\beta_i v^\beta_{LM} \ Y_{LM} (\hat{n}_i)   =  \sum_{\beta=0}^{N-1} \tau^\alpha_i \bar{\tau}^\beta_i v^\beta_i.
\end{equation}

\section{Coarse-graining}\label{app:coarsegraining_harmonic}

To determine if there is a solution to the maximum likelihood condition Eq.~\eqref{eq:appmaxLcond}, we must determine if the operator $\boldsymbol{\tau}^\dagger (\boldsymbol{C}^{\mathrm{pCMB}})^{-1} \boldsymbol{\tau}$ is invertible for a particular coarse graining procedure. Note that this operator depends on the {\em actual realization} of the optical depth anisotropies, so there is some danger that an inverse exists for some realizations but not others. Nevertheless, we can get some idea of what to expect by examining the statistical properties of this operator over an ensemble of realizations of the optical depth anisotropies. For example, if we focus on the harmonic basis, the mean of the operator is:
 \begin{eqnarray}\label{eq:matrixmean}
\langle \left[ \boldsymbol{\tau}^\dagger (\boldsymbol{C}^{\mathrm{pCMB}})^{-1} \boldsymbol{\tau} \right]^{\alpha \beta}_{L'M' ; LM} \rangle_{\tau} &=& \sum_{\ell, \ell'} \frac{(2 \ell + 1)(2 \ell'+1)}{4 \pi} 
\left(\begin{array}{ccc}
\ell & \ell' & L \\
0 & 0 & 0
\end{array}\right)^2
\frac{C_{\ell'}^{\tau^\alpha \tau^\beta}}{C_\ell^{\mathrm{pCMB}}} \delta_{LL'} \delta_{MM'}
\end{eqnarray}
where the $\langle \ldots \rangle_{\tau}$ denotes an ensemble average over the optical depth anisotropies. On average, we see that the operator is diagonal in harmonic space. If the sum is dominated by large-$\ell$ (which we expect it to be) then it will also be nearly diagonal in bin-space. This is because the bin-bin correlations in $C_\ell^{\tau^a \tau^b}$ become vanishingly small at high-$\ell$ as the scales probed by the angular projection become much smaller than the extent of the bins. Finding the variance of the operator:
 \begin{eqnarray}
\langle \left( \left[\boldsymbol{\tau}^\dagger (\boldsymbol{C}^{\mathrm{pCMB}})^{-1} \boldsymbol{\tau} \right]^{\alpha \beta}_{L'M' ; LM} \right)^2 \rangle 
&\simeq&  \langle \left[\boldsymbol{\tau}^\dagger (\boldsymbol{C}^{\mathrm{pCMB}})^{-1} \boldsymbol{\tau} \right]^{\alpha \beta}_{L'M' ; LM} \rangle^2 \nonumber \\
&+&  \left( \sum_\ell \frac{2\ell+1}{4\pi} \left[ \frac{C_\ell^{\tau^\alpha \tau^\alpha} C_{\ell}^{\tau^\beta \tau^\beta}}{(C_{\ell}^{\mathrm{pCMB}})^2 } + \frac{C_\ell^{\tau^\alpha \tau^\beta} C_{\ell}^{\tau^\alpha \tau^\beta}}{(C_{\ell}^{\mathrm{pCMB}})^2 }    \right] \right) \sum_{\ell''} (2 \ell''+1) (2L+1) (2L'+1) \nonumber \\ 
&\times&  
\left(\begin{array}{ccc}
L & L' & \ell'' \\
0 & 0 & 0
\end{array}\right)^2
\left(\begin{array}{ccc}
L & L' & \ell'' \\
M & M' & M-M'
\end{array}\right)^2 
\end{eqnarray}
where we have assumed that the optical depth anisotropies are Gaussian and used the limit where $L,L'$ are small. Notably, the variance is not diagonal in the harmonic coefficients, and becomes increasingly non-diagonal at higher $L$ and $L'$. It is also non-diagonal in bin space. Therefore, in a particular realization we can expect that the operator $\boldsymbol{\tau}^\dagger (\boldsymbol{C}^{\mathrm{pCMB}})^{-1} \boldsymbol{\tau} $ becomes highly non-diagonal on small angular scales. This implies that the condition number could be quite large at full resolution, and there may be no well-defined inverse. Our strategy is therefore to coarse-grain the maximum likelihood condition by truncating at some empirically determined $L_{\rm max}$ and $N$ before solving for the velocity field.

 \section{Connecting to the quadratic estimator}\label{sec:connectQE}
 
 \subsection{Harmonic space estimator}
 
In the low signal-to-noise regime, we can obtain the quadratic estimator from the maximum likelihood condition Eq.~\eqref{eq:harmonicest} by taking an ensemble average over the optical depth $\langle \rangle_\tau$:
\begin{equation}
\left[ \boldsymbol{\tau}^\dagger (\boldsymbol{C}^{\mathrm{pCMB}})^{-1} \Theta \right]_{L'M'}^\alpha = \sum_{LM; \ \beta} \langle \left[\boldsymbol{ \tau}^\dagger (\boldsymbol{C}^{\mathrm{pCMB}})^{-1} \boldsymbol{\tau} \right]^{\alpha \beta}_{L'M' ; LM} \rangle_\tau \ v^\beta_{LM}.
\end{equation}
The ensemble average is given by Eq.~\eqref{eq:matrixmean}, which using the notation of Ref.~\cite{Cayuso:2021ljq} can be recognized as:
\begin{equation}
\langle \left[ \boldsymbol{\tau}^\dagger (\boldsymbol{C}^{\mathrm{pCMB}})^{-1} \boldsymbol{\tau} \right]^{\alpha \beta}_{L'M' ; LM} \rangle_\tau = \frac{1}{2L+1} \sum_{\ell, \ell'} G^{v^\beta \tau^\beta}_{\ell \ell' L} f^{v^\beta \tau^\alpha}_{\ell L \ell'} \delta_{LL'} \delta_{MM'}
\end{equation}
with
\begin{equation}
G^{v^\beta \tau^\beta}_{\ell \ell' L} = \frac{f_{\ell L \ell'}^{v^\beta \tau^\beta} }{C_\ell^{\mathrm{pCMB}} C_{\ell'}^{\tau^\beta \tau^\beta} };\ \ \ f_{\ell L \ell'}^{v^\beta \tau^\alpha} = \sqrt{\frac{(2\ell+1) (2\ell'+1) (2L+1)}{4 \pi}} \left(\begin{array}{ccc}
\ell & \ell' & L \\
0 & 0 & 0
\end{array}\right)  C_{\ell'}^{\tau^\beta \tau^\alpha}.
\end{equation}
The left hand side of the maximum likelihood condition can be expanded to: 
\begin{equation}
\left[ \boldsymbol{\tau}^\dagger (\boldsymbol{C}^{\mathrm{pCMB}})^{-1} \Theta \right]_{L M}^\alpha = \sum_{\ell m; \ell' m'} (-1)^{M} 
\left(\begin{array}{ccc}
\ell & \ell' & L \\
m & m' & -M
\end{array}\right) G^{v^\alpha \tau^\alpha}_{\ell \ell' L} \Theta_{\ell m} \tau^\alpha_{\ell' m'}.
\end{equation}
For a completely diagonal $C_{\ell'}^{\tau^\beta \tau^\alpha} = C_{\ell'}^{\tau^\alpha \tau^\alpha} \delta_{\alpha \beta}$, we solve the ensemble-averaged maximum likelihood condition to arrive at the traditional form for the quadratic estimator appearing in Refs.~\cite{Deutsch:2017ybc,Cayuso:2021ljq}:
\begin{equation}\label{eq:harmonicQE_end}
v^\alpha_{LM} = A_L^{v^\alpha} \sum_{\ell m; \ell' m'} (-1)^{M} 
\left(\begin{array}{ccc}
\ell & \ell' & L \\
m & m' & -M
\end{array}\right) G^{v^\alpha \tau^\alpha}_{\ell \ell' L} \Theta_{\ell m} \tau^\alpha_{\ell' m'}
\end{equation}
where 
\begin{equation}\label{eq:ALalpha}
A_L^{v^\alpha} \equiv \left[ \frac{1}{2L+1} \sum_{\ell, \ell'} G^{v^\alpha \tau^\alpha}_{\ell \ell' L} f^{v^\alpha \tau^\alpha}_{\ell L \ell'}  \right]^{-1}.
\end{equation}
There are, however, a few crucial differences with the quadratic estimator presented in Refs.~\cite{Deutsch:2017ybc,Cayuso:2021ljq}. In the traditional quadratic estimator, the weights needed for an unbiased minimum variance reconstruction require that everywhere above we replace $C_\ell^{\mathrm{pCMB}} \rightarrow C_\ell^{\mathrm{pCMB}} + C_\ell^{\rm{kSZ}}$. In the limit where $C_\ell^{\mathrm{pCMB}} \ll C_\ell^{\rm{kSZ}}$, as occurs for a sufficiently high-resolution low-noise CMB experiment, this is a non-negligible difference -- as expected, since in this case the temperature anisotropies on small scales are very non-Gaussian, and far from the limit of Gaussian fields used to derive the quadratic estimator. Therefore, we only recover the traditional quadratic estimator in the limit where $C_\ell^{\mathrm{pCMB}} \gg C_\ell^{\rm{kSZ}}$. This observation supports the notion that the maximum likelihood estimator will be superior in the high signal-to-noise regime. Another difference arises in the case where $C_{\ell}^{\tau^\beta \tau^\alpha}$ is not diagonal. In this case, to find a solution to the ensemble averaged maximum likelihood condition it is necessary to find the inverse of $C_{\ell}^{\tau^\beta \tau^\alpha}$ (if it exists). The resulting solution is different than the approach presented in Ref.~\cite{Cayuso:2021ljq}, which first constructs a set of biased estimators which are then de-biased by forming the appropriate linear combinations. The relevant differences would be interesting to explore further in future work.

\subsection{Map space estimator}\label{sec:mapQE}

Just as for the harmonic space estimator, we can derive an ensemble average map space estimator. Starting from Eq.~\eqref{eq:pixelML}: 
\begin{eqnarray}
\int d\Omega \ W_I (\hat{n}) \tau^\alpha (\hat{n}) \bar{\Theta} (\hat{n}) &=& \sum_{\beta=0}^{N-1} v^\beta_I \ \int d\Omega \ W_I (\hat{n})  \langle \tau^\alpha (\hat{n}) \bar{\tau}^\beta (\hat{n}) \rangle_\tau \\ &+&  \sum_{\beta=0}^{N-1} \int d\Omega \ W_I (\hat{n})\langle \tau^\alpha (\hat{n}) \bar{\tau}^\beta (\hat{n})  \rangle_\tau  \Delta v^\beta_I (\hat n) \\
&=&  \sum_{\beta=0}^{N-1} v^\beta_I\ \langle \tau^\alpha (0) \bar{\tau}^\beta (0) \rangle_\tau \ \int d\Omega \ \left[ W_I (\hat{n}) + \int d\Omega \ W_I (\hat{n}) \Delta v^\beta_I (\hat n)  \right] \\
&=&  \sum_{\beta=0}^{N-1} v^\beta_I \ \langle \tau^\alpha (0) \bar{\tau}^\beta (0) \rangle_\tau
\end{eqnarray}
where we've assumed that $\tau^\alpha$ is statistically isotropic, $\int d\Omega \ W_I (\hat{n}) = 1$, and $\int d\Omega \ W_I (\hat{n}) \Delta v^\beta_I (\hat n) = 0$ by definition. This suggests the following map space quadratic estimator:
\begin{equation}\label{eq:mapsimpleQE}
\hat{v}^\alpha_I = \sum_{\beta=0}^{N-1} \left[ \langle \tau^\beta (0) \bar{\tau}^\alpha (0) \rangle_\tau \right]^{-1} \int d\Omega \ W_I (\hat{n}) \tau^\alpha (\hat{n}) \bar{\Theta} (\hat{n}).
\end{equation}
For a diagonal variance $\langle \tau^\beta (0) \bar{\tau}^\alpha (0) \rangle_\tau = \langle \tau^\alpha (0) \bar{\tau}^\alpha (0) \rangle_\tau \delta_{\alpha \beta}$, we can identify
\begin{equation}
\langle \tau^\alpha (0) \bar{\tau}^\alpha (0) \rangle_\tau = \frac{1}{A_{L=0}^{v^\alpha}}
\end{equation}
as defined above in Eq.~\eqref{eq:ALalpha}. Since Eq.~\eqref{eq:ALalpha} is nearly independent of $L$~\cite{Deutsch:2017ybc,Cayuso:2021ljq}, one can simply show that Eq.~\eqref{eq:mapsimpleQE} is equivalent to Eq.~\eqref{eq:harmonicQE_end} at low $L$, and also equivalent to the map based quadratic estimator presented in Refs.~\cite{Deutsch:2017ybc,Cayuso:2021ljq} (with the same caveats as for the harmonic space estimator). In the low signal-to-noise regime, Eq.~\eqref{eq:mapsimpleQE} may be a simpler implementation of the quadratic estimator used in previous work. We leave further investigation of this point to future work.

\section{Mean and variance of the MaxL estimator}\label{app:noise_covariance}

\subsection{Harmonic space estimator}\label{app:noise_covariance_harmonic}

The mean of the harmonic space estimator is
\begin{equation}
\langle \hat{v}_{LM}^\alpha \rangle_{\Theta} =  \sum_{L'M'}^{L'_{\rm{max}}} \sum_{\beta=0}^{N-1} \left( [\boldsymbol{\tau}^\dagger (\boldsymbol{C}^{\mathrm{pCMB}})^{-1} \boldsymbol{\tau}]^{-1} \right)^{\alpha \beta}_{LM ; L'M'}  \langle \left[\boldsymbol{\tau}^\dagger (C^{\mathrm{pCMB}})^{-1} \Theta \right]_{L'M'}^\beta \rangle_{\Theta}.
\end{equation}

To calculate this, we must evaluate:
\begin{eqnarray}
 \langle \left[ \boldsymbol{\tau}^\dagger (\boldsymbol{C}^{\mathrm{pCMB}})^{-1} \Theta \right]_{L'M'}^\alpha \rangle_{\Theta} &=& \int d\Omega \ \tau^\alpha (\hat{n}) \ \langle \bar{\Theta} (\hat{n}) \rangle_{\Theta} \ Y_{L'M'} ^*(\hat{n}) .
\end{eqnarray}
The observed temperature anisotropies consist of the zero-mean primary CMB and the kSZ:
\begin{eqnarray}
\langle \bar{\Theta} (\hat{n}) \rangle_{\Theta} = \langle \bar{\Theta}^{\mathrm{pCMB}} (\hat{n}) \rangle_{\Theta} + \bar{\Theta}^{\rm{kSZ}} (\hat{n}) = \bar{\Theta}^{\rm{kSZ}} (\hat{n})
\end{eqnarray}
Expanding the kSZ anisotropies into coarse- and fine-grained components, we have
\begin{eqnarray}
 \langle \left[ \boldsymbol{\tau}^\dagger (\boldsymbol{C}^{\mathrm{pCMB}})^{-1} \Theta \right]_{L'M'}^\alpha \rangle_{\Theta} &=& \int d\Omega \ \tau^\alpha (\hat{n}) \ \left[ \sum_{\beta=0}^{N-1} \sum_{LM}^{L_{\rm{max}}} \bar{\tau}^\beta (\hat{n}) v^\beta_{LM} Y_{LM} (\hat{n}) \right. \nonumber \\
 &&\left. + \sum_{b = N}^{\infty} \sum_{LM}^{L_{\rm{max}}} \bar{\tau}^b (\hat{n}) v^b_{LM} Y_{LM} (\hat{n}) + \sum_{b=0}^\infty \sum_{LM=L_{max}}^{\infty} \bar{\tau}^b (\hat{n}) v^b_{LM} Y_{LM} (\hat{n}) \right] \ Y_{L'M'} ^*(\hat{n}) \nonumber \\
 &\simeq&  \sum_{\beta=0}^{N-1} \sum_{LM}^{L_{\rm{max}}} [\boldsymbol{\tau}^\dagger (\boldsymbol{C}^{\mathrm{pCMB}})^{-1} \boldsymbol{\tau}]^{\alpha \beta}_{L'M' ; LM} v_{LM}^\beta + \sum_{b=N}^\infty \sum_{LM}^{L_{\rm{max}}} [\boldsymbol{\tau}^\dagger (\boldsymbol{C}^{\mathrm{pCMB}})^{-1} \boldsymbol{\tau}]^{\alpha b}_{L'M' ; LM} v_{LM}^b \nonumber
\end{eqnarray}
where we assumed that the dominant contributions from the velocity field are at low $L$. We therefore have
\begin{equation}
\langle \hat{v}_{LM}^\alpha \rangle_{\Theta} = v_{LM}^\alpha +  \beta_{LM}^\alpha
\end{equation}
where
\begin{equation}
\beta_{LM}^\alpha \equiv \sum_{L'M'}^{L_{\rm{max}}} \sum_{\gamma = 0}^{N-1} \sum_{c=N}^\infty \sum_{L''M''}^{L_{\rm{max}}}  \left( [\boldsymbol{\tau}^\dagger (\boldsymbol{C}^{\mathrm{pCMB}})^{-1} \boldsymbol{\tau}]^{-1} \right)^{\alpha \gamma}_{LM ; L'M'} [\boldsymbol{\tau}^\dagger (\boldsymbol{C}^{\mathrm{pCMB}})^{-1} \boldsymbol{\tau}]^{\gamma c}_{L'M' ; L''M''} v_{L''M''}^c.
\end{equation}
So, the estimator is unbiased up to an additive bias $\beta_{LM}^\alpha$, which depends on the fine-grained radial velocity modes.

Moving to the variance:
\begin{eqnarray}
\langle \hat{v}^\alpha_{LM} \hat{v}^\beta_{L'M'}  \rangle_{\Theta} &=& 
 \sum_{L_a M_a; L_b M_b}^{L_{\rm max}} \sum_{\gamma; \delta=0}^{N-1}  \left( [\boldsymbol{\tau}^\dagger (\boldsymbol{C}^{\mathrm{pCMB}})^{-1} \boldsymbol{\tau}]^{-1} \right)^{\alpha \gamma}_{LM ; L_a M_a} \left( [\boldsymbol{\tau}^\dagger (\boldsymbol{C}^{\mathrm{pCMB}})^{-1} \boldsymbol{\tau}]^{-1} \right)^{\beta \delta}_{L'M' ; L_b M_b}  \nonumber \\
  &\times& \int \int d\Omega d\Omega' \ \tau^\gamma (\hat{n}) \tau^\delta (\hat{n}') \ \langle \bar{\Theta}(\hat{n}) \bar{\Theta}(\hat{n}') \rangle_\Theta \ Y_{L_a M_a} (\hat{n}) Y_{L_b M_b} (\hat{n}').
\end{eqnarray}
There is a contribution to the temperature correlation function from the primary CMB and from the kSZ effect:
\begin{equation}
\langle \bar{\Theta}(\hat{n}) \bar{\Theta}(\hat{n}') \rangle_\Theta = \sum_{\ell m} \frac{1}{C_{\ell}^{\mathrm{pCMB}}} Y_{\ell m} (\hat{n}) Y_{\ell m} (\hat{n}') + \bar{\Theta}^{\rm{kSZ}} (\hat{n}) \bar{\Theta}^{\rm{kSZ}} (\hat{n}').
\end{equation}
The first term will give rise to the Gaussian estimator noise and the second will describe our signal and biases arising due to our coarse-graining of the maximum likelihood condition:
\begin{equation}
\langle \hat{v}^\alpha_{LM} \hat{v}^\beta_{L'M'}  \rangle_{\Theta} = (C^{vv})_{LM; L'M'}^{\alpha \beta} + N_{LM; L'M'}^{\alpha \beta}.
\end{equation}
The reconstruction noise $N_{LM; L'M'}^{\alpha \beta}$ arises due to the contribution from the primary CMB:
\begin{eqnarray}
N_{LM; L'M'}^{\alpha \beta} &=& \sum_{L_a M_a; L_b M_b; \ell m} \sum_{\gamma; \delta}  \left( [\boldsymbol{\tau}^\dagger (\boldsymbol{C}^{\mathrm{pCMB}})^{-1} \boldsymbol{\tau}]^{-1} \right)^{\alpha \gamma}_{LM ; L_a M_a} \left( [\boldsymbol{\tau}^\dagger (\boldsymbol{C}^{\mathrm{pCMB}})^{-1}\boldsymbol{ \tau}]^{-1} \right)^{\beta \delta}_{L'M' ; L_b M_b}  \nonumber \\
  &\times& \int \int d\Omega d\Omega' \ \tau^\gamma (\hat{n}) \tau^\delta (\hat{n}') \frac{1}{C_{\ell}^{\mathrm{pCMB}}} \ Y_{\ell m} (\hat{n}) Y_{\ell m} (\hat{n}') Y_{L_a M_a} (\hat{n}) Y_{L_b M_b} (\hat{n}') \\
  &\simeq& \sum_{L_a M_a; L_b M_b} \sum_{\gamma; \delta}  \left( [\boldsymbol{\tau}^\dagger (\boldsymbol{C}^{\mathrm{pCMB}})^{-1} \boldsymbol{\tau}]^{-1} \right)^{\alpha \gamma}_{LM ; L_a M_a} \left( [\boldsymbol{\tau}^\dagger (\boldsymbol{C}^{\mathrm{pCMB}})^{-1} \boldsymbol{\tau}]^{-1} \right)^{\beta \delta}_{L'M' ; L_b M_b}  \nonumber \\
  &\times& \int d\Omega \ \tau^\gamma (\hat{n}) \bar{\tau}^\delta (\hat{n}) \ Y_{L_a M_a} (\hat{n}) Y_{L_b M_b} (\hat{n}) \\
  &=& \sum_{L_a M_a; L_b M_b} \sum_{\gamma; \delta}  \left( [\boldsymbol{\tau}^\dagger (\boldsymbol{C}^{\mathrm{pCMB}})^{-1} \boldsymbol{\tau}]^{-1} \right)^{\alpha \gamma}_{LM ; L_a M_a} \left( [\boldsymbol{\tau}^\dagger (\boldsymbol{C}^{\mathrm{pCMB}})^{-1} \boldsymbol{\tau}]^{-1} \right)^{\beta \delta}_{L'M' ; L_b M_b} \\
    &\times&  \left( \boldsymbol{\tau}^\dagger (\boldsymbol{C}^{\mathrm{pCMB}})^{-1} \boldsymbol{\tau} \right)^{\gamma \delta}_{L_b M_b ; L_a M_a} \\
      &=&  \left( [\boldsymbol{\tau}^\dagger (\boldsymbol{C}^{\mathrm{pCMB}})^{-1} \boldsymbol{\tau}]^{-1} \right)^{\alpha \beta}_{LM ; L'M'} 
\end{eqnarray}
where in going from the first to second lines we have assumed that $C_\ell^{\mathrm{pCMB}}$ does not vary too quickly on scales of order $\Delta \ell = L$ and used the identity
\begin{equation}
\int d^2 n \ Y_{\ell_a m_a} Y_{\ell_b m_b} Y_{\ell_c m_c} Y_{\ell_d m_d} =  \sum_{\ell m} \left( \int d^2 n \ Y_{\ell_a m_a} Y_{\ell_b m_b} Y_{\ell m} \right) \left(\int d^2 n' \ Y_{\ell_c m_c} Y_{\ell_d m_d} Y_{\ell m}\right).
\end{equation}

For the signal and bias due to `fine modes' we have: 
\begin{eqnarray}
(C^{vv})_{LM; L'M'}^{\alpha \beta} &=&  \sum_{L_a M_a; L_b M_b} \sum_{\gamma; \delta}  \left( [\boldsymbol{\tau}^\dagger (\boldsymbol{C}^{\mathrm{pCMB}})^{-1} \boldsymbol{\tau}]^{-1} \right)^{\alpha \gamma}_{LM ; L_a M_a} \left( [\boldsymbol{\tau}^\dagger (\boldsymbol{C}^{\mathrm{pCMB}})^{-1} \boldsymbol{\tau}]^{-1} \right)^{\beta \delta}_{L'M' ; L_b M_b}  \nonumber \\
  &\times& \int \int d\Omega d\Omega' \ \tau^\gamma (\hat{n}) \tau^\delta (\hat{n}') \ \bar{\Theta}^{\rm{kSZ}} (\hat{n}) \bar{\Theta}^{\rm{kSZ}} (\hat{n}') \ Y_{L_a M_a} (\hat{n}) Y_{L_b M_b} (\hat{n}')  \\
  &\simeq&   \sum_{L_a M_a; L_b M_b} \sum_{\gamma; \delta; e; f}  \left( [\boldsymbol{\tau}^\dagger (\boldsymbol{C}^{\mathrm{pCMB}})^{-1} \boldsymbol{\tau}]^{-1} \right)^{\alpha \gamma}_{LM ; L_a M_a} \left( [\boldsymbol{\tau}^\dagger (\boldsymbol{C}^{\mathrm{pCMB}})^{-1} \boldsymbol{\tau}]^{-1} \right)^{\beta \delta}_{L'M' ; L_b M_b}  \nonumber \\
  &\times& \int \int d\Omega d\Omega' \ \tau^\gamma (\hat{n}) \tau^\delta (\hat{n}') \bar{\tau}^e (\hat{n})  v^e (\hat{n})  \bar{\tau}^f (\hat{n}') v^f (\hat{n}') \ Y_{L_a M_a} (\hat{n}) Y_{L_b M_b} (\hat{n}') \nonumber \\
  &=&  \sum_{L_a M_a; L_b M_b}\sum_{\gamma; \delta; e; f} \left( [\boldsymbol{\tau}^\dagger (\boldsymbol{C}^{\mathrm{pCMB}})^{-1} \boldsymbol{\tau}]^{-1} \right)^{\alpha \gamma}_{LM ; L_a M_a} \left( [\boldsymbol{\tau}^\dagger (\boldsymbol{C}^{\mathrm{pCMB}})^{-1} \boldsymbol{\tau}]^{-1} \right)^{\beta \delta}_{L'M' ; L_b M_b}  \nonumber \\
  &\times& \sum_{L_c M_c; L_d M_d} v_{L_c M_c}^e v_{L_d M_d}^f  \int d\Omega   \ \tau^\gamma (\hat{n}) \bar{\tau}^e (\hat{n}) Y_{L_a M_a} (\hat{n}) Y_{L_c M_c} (\hat{n}) \int d\Omega'  \tau^\delta (\hat{n}') \bar{\tau}^f (\hat{n}') Y_{L_b M_b} (\hat{n}') Y_{L_d M_d} (\hat{n}')  \nonumber \\
  &=& \sum_{L_a M_a; L_b M_b}  \sum_{L_c M_c; L_d M_d} \sum_{\gamma; \delta; e; f} v_{L_c M_c}^e v_{L_d M_d}^f  \left( [\boldsymbol{\tau}^\dagger (\boldsymbol{C}^{\mathrm{pCMB}})^{-1} \boldsymbol{\tau}]^{-1} \right)^{\alpha \gamma}_{LM ; L_a M_a} \left[ \boldsymbol{\tau}^\dagger (\boldsymbol{C}^{\mathrm{pCMB}})^{-1} \boldsymbol{\tau} \right]^{\gamma e}_{L_a M_a ; L_c M_c}    \nonumber \\ 
  &\times& \left( [\boldsymbol{\tau}^\dagger (\boldsymbol{C}^{\mathrm{pCMB}})^{-1} \boldsymbol{\tau}]^{-1} \right)^{\beta \delta}_{L'M' ; L_b M_b} \left[ \boldsymbol{\tau}^\dagger (\boldsymbol{C}^{\mathrm{pCMB}})^{-1} \boldsymbol{\tau} \right]^{\delta f}_{L_b M_b ; L_d M_d}  \nonumber \\
   &=& (v_{L M}^\alpha + \beta_{L M}^\alpha) (v_{L' M'}^\beta +  \beta_{L' M'}^\beta).
 \end{eqnarray}
The main assumption made in the derivation of the estimator mean and variance was that $C_\ell^{\mathrm{pCMB}}$ does not vary too quickly on scales of order $\Delta \ell = L$, which can always be satisfied for sufficiently small $L$. For modest values of $L \sim 100$, we expect this to be a reasonable assumption at large $\ell$ where the primary CMB is smooth and dominated by lensing, instrumental noise, or reionization kSZ (depending on the instrument).

\subsection{Map space estimator}\label{app:noise_covariance_map}

We now compute the mean and variance of the map space estimator. Starting with the mean:
\begin{eqnarray}
\langle \hat{v}_I^\alpha \rangle_{\Theta} &=&  \sum_{\beta = 0}^{N-1} \Wop{I}{\alpha}{\beta}{}^{-1} \int d\Omega \ W_I (\hat{n}) \tau^\beta (\hat{n}) \bar{\Theta}^{\rm kSZ} (\hat{n}), \\
&=&  \sum_{\beta = 0}^{N-1} \Wop{I}{\alpha}{\beta}{}^{-1} \int d\Omega \ W_I (\hat{n}) \tau^\beta (\hat{n}) \sum_{c = 0}^{\infty} \bar{\tau}^c (\hat{n}) \left( v^c_I + \Delta v^c_I (\hat{n})  \right)  \\
&=&  \sum_{\beta; \gamma = 0}^{N-1} v^\gamma_I \Wop{I}{\alpha}{\beta}{}^{-1} \ \Wop{I}{\gamma}{\beta} + \sum_{\beta; \gamma = 0}^{N-1} \Wop{I}{\alpha}{\beta}{}^{-1} \int d\Omega \ W_I (\hat{n}) \tau^\beta (\hat{n}) \bar{\tau}^\gamma (\hat{n})  \Delta v_I^\gamma (\hat{n}) \\
&+& \sum_{\beta=0}^{N-1} \sum_{c = N}^{\infty} \Wop{I}{\alpha}{\beta}{}^{-1} \int d\Omega \ W_I (\hat{n}) \tau^\beta (\hat{n}) \bar{\tau}^c (\hat{n})  v^c (\hat{n}) \\
&=& v_I^\alpha + \beta_I^\alpha 
\end{eqnarray}
where we have defined 
\begin{eqnarray}
\beta_I^\alpha &=& \sum_{\beta; \gamma = 0}^{N-1} \Wop{I}{\alpha}{\beta}{}^{-1} \int d\Omega \ W_I (\hat{n}) \tau^\beta (\hat{n}) \bar{\tau}^\gamma (\hat{n}) \Delta v^\gamma_I (\hat{n}) \nonumber \\ 
&+&\sum_{\beta=0}^{N-1} \sum_{c=N}^\infty \Wop{I}{\alpha}{\beta}{}^{-1}\int d\Omega \ W_I (\hat{n}) \tau^\beta (\hat{n}) \bar{\tau}^c (\hat{n}) v^c (\hat{n}). 
\end{eqnarray}

Moving to the estimator variance, we have:
\begin{eqnarray}
\langle \hat{v}_I^\alpha \hat{v}_J^\beta \rangle_{\Theta} &=& \sum_{\gamma;\delta = 0}^{N-1}\Wop{I}{\alpha}{\gamma}{}^{-1} \Wop{J}{\beta}{\delta}{}^{-1}  \int d\Omega d\Omega' \ W_I(\hat{n}) W_J (\hat{n}') \tau^\gamma (\hat{n}) \tau^\delta (\hat{n}')     \langle \bar{\Theta} (\hat{n}) \bar{\Theta} (\hat{n}') \rangle_{\Theta} \nonumber \\
&=& \sum_{\gamma;\delta = 0}^{N-1}\Wop{I}{\alpha}{\gamma}{}^{-1} \Wop{J}{\beta}{\delta}{}^{-1} \left[   \int d\Omega \ W_I(\hat{n}) \tau^\gamma (\hat{n}) \bar{\Theta}^{\rm{kSZ}} (\hat{n}) \int d\Omega' \ W_J (\hat{n}')  \tau^\delta (\hat{n}')   \bar{\Theta}^{\rm{kSZ}} (\hat{n}') \right. \nonumber \\
&+& \left.  \sum_{\ell m} \int d\Omega \ W_I(\hat{n}) \tau^\gamma (\hat{n}) Y_{\ell m} (\hat{n})  \int  d\Omega' \ W_J (\hat{n}') \frac{\tau^\delta (\hat{n}') }{C_{\ell}^{\mathrm{pCMB}}} Y_{\ell m} (\hat{n}') \right]  \nonumber \\
&\simeq& \left( v_I^\alpha + \beta_I^\alpha  \right) \left( v_J^\beta + \beta_J^\beta \right) +  \sum_{\gamma ;\delta = 0}^{N-1} \Wop{I}{\alpha}{\gamma}{}^{-1}\Wop{J}{\beta}{\delta}{}^{-1} \int d\Omega \ W_I(\hat{n}) W_J (\hat{n}) \tau^\gamma (\hat{n}) \bar{\tau}^\delta (\hat{n}) \nonumber \\
&=&\left( v_I^\alpha + \beta_I^\alpha  \right) \left( v_J^\beta + \beta_J^\beta \right) + \frac{1}{\Delta\theta^2} \Wop{I}{\alpha}{\beta}{}^{-1}\delta_{IJ}
\end{eqnarray}
where in going from the second to the third line, we have made the approximation that $C_{\ell}^{\mathrm{pCMB}}$ is not too rapidly varying with scale, as we did in the derivation of the harmonic space estimator variance.

\section{Optical depth bias}\label{sec:opticaldepthbias}

When we only have access to an estimate $\hat{\tau}$ of the differential optical depth, the harmonic space estimator is given by
\begin{equation}
\hat{v}_{LM}^\alpha = \sum_{L'M'}^{L'_{\rm{max}}} \sum_{\beta=0}^{N-1} \left( [\boldsymbol{\hat{\tau}}^\dagger (\boldsymbol{C}^{\mathrm{pCMB}})^{-1} \boldsymbol{\hat{\tau}}]^{-1} \right)^{\alpha \beta}_{LM ; L'M'}  (\boldsymbol{\hat{\tau}}^\dagger (\boldsymbol{C}^{\mathrm{pCMB}})^{-1} \Theta)_{L'M'}^\beta
\end{equation}
Computing the estimator mean, we have:
\begin{eqnarray}
\langle \hat{v}_{LM}^\alpha \rangle_{\Theta} &=&  \sum_{L'M'}^{L'_{\rm{max}}} \sum_{\beta=0}^{N-1} \left( [ \boldsymbol{\hat{\tau}}^\dagger (\boldsymbol{C}^{\mathrm{pCMB}})^{-1} \boldsymbol{\hat{\tau}}]^{-1} \right)^{\alpha \beta}_{LM ; L'M'}  \langle (\boldsymbol{\hat{\tau}}^\dagger (\boldsymbol{C}^{\mathrm{pCMB}})^{-1} \Theta)_{L'M'}^\beta \rangle_{\Theta} \\
&=& \left[ \sum_{L''M''}^{L_{\rm{max}}} \sum_{\gamma=0}^{N-1}  R^{\alpha \gamma}_{LM;L''M''} v_{L''M''}^\gamma \right]+ \hat{\beta}^\alpha_{LM}
\end{eqnarray}
where
\begin{equation}
R^{\alpha \gamma}_{LM;L''M''} = \sum_{L'M'}^{L'_{\rm{max}}} \sum_{\beta=0}^{N-1}  \left( [\boldsymbol{ \hat{\tau}}^\dagger (\boldsymbol{C}^{\mathrm{pCMB}})^{-1} \boldsymbol{\hat{\tau}}]^{-1} \right)^{\alpha \beta}_{LM ; L'M'}  [\boldsymbol{\hat{\tau}}^\dagger (\boldsymbol{C}^{\mathrm{pCMB}})^{-1} \boldsymbol{\tau}]^{\beta \gamma}_{L'M' ; L''M''}
\end{equation}
and
\begin{equation}
\hat{\beta}_{LM}^\alpha \equiv \sum_{L'M'}^{L_{\rm{max}}} \sum_{\gamma = 0}^{N-1} \sum_{c=N}^\infty \sum_{L''M''}^{L_{\rm{max}}}  \left( [\boldsymbol{\hat{\tau}}^\dagger (\boldsymbol{C}^{\mathrm{pCMB}})^{-1} \boldsymbol{\hat{\tau}}]^{-1} \right)^{\alpha \gamma}_{LM ; L'M'} [\boldsymbol{\hat{\tau}}^\dagger (\boldsymbol{C}^{\mathrm{pCMB}})^{-1} \boldsymbol{\tau}]^{\gamma c}_{L'M' ; L''M''} v_{L''M''}^c.
\end{equation}
When $\hat{\tau} = \tau$, this is simply the identity matrix, and we recover an unbiased estimator mean. In reality, an imperfect estimate of the differential optical depth will result in the mixing of information and therefore a bias. It is important to characterize the magnitude and properties of this bias to obtain useful cosmological information from the reconstruction. 

To explore the optical depth bias a bit more, suppose we have a pretty good reconstruction of the differential optical depth, such that
\begin{equation}\label{eq:tausnmodel}
\hat{\tau}^\alpha (\hat{n}) = \tau^\alpha (\hat{n}) + n_{\tau}^\alpha  (\hat{n}); \ \ \ n_{\tau}^\alpha  (\hat{n}) \ll \tau^\alpha (\hat{n}).
\end{equation}
We can approximate 
\begin{equation}
[ \boldsymbol{\hat{\tau}}^\dagger (\boldsymbol{C}^{\mathrm{pCMB}})^{-1} \boldsymbol{\hat{\tau}}]^{-1}  \simeq \left[ 1 - [\boldsymbol{\tau}^\dagger (\boldsymbol{C}^{\mathrm{pCMB}})^{-1}\boldsymbol{ \tau}]^{-1}  \cdot [\boldsymbol{n_{\tau}}^\dagger (\boldsymbol{C}^{\mathrm{pCMB}})^{-1} \boldsymbol{n_{\tau}}]  \right] \cdot [\boldsymbol{\tau}^\dagger (\boldsymbol{C}^{\mathrm{pCMB}})^{-1} \tau]^{-1} 
\end{equation}
and therefore
\begin{equation}
R^{\alpha \gamma}_{LM;L''M''} \simeq \delta_{\alpha \gamma} \delta_{LM; L''M''} - \sum_{L'M'}^{L'_{\mathrm{max}}} \sum_{\beta=0}^{N-1} \left([\boldsymbol{\tau}^\dagger (\boldsymbol{C}^{\mathrm{pCMB}})^{-1} \boldsymbol{\tau}]^{-1}\right)^{\alpha \beta}_{LM; L'M'}  [\boldsymbol{n_{\tau}}^\dagger (\boldsymbol{C}^{\mathrm{pCMB}})^{-1} \boldsymbol{n_{\tau}}]^{\beta \gamma}_{L'M'; L''M''} .
\end{equation}
Taking an ensemble average over $\tau$ and $n_{\tau}$, and using Eq.~\eqref{eq:matrixmean}, we have 
\begin{equation}
\langle R^{\alpha \gamma}_{LM;L''M''} \rangle \simeq \delta_{\alpha \gamma} \delta_{LM; L''M''} - \delta_{LM; L''M''} \sum_{\beta}^{\beta_{\mathrm{max}}} \langle \tau^\alpha (0) \bar{\tau}^\beta (0)\rangle^{-1} \langle n_{\tau}^\beta (0) \bar{n}_{\tau}^\gamma (0) \rangle
\end{equation}
where $\langle \tau^\alpha (0) \bar{\tau}^\beta (0)\rangle^{-1}$ is the bin-space inverse of Eq.~\eqref{eq:matrixmean} and $\langle n_{\tau}^\beta (0) \bar{n}_{\tau}^\gamma (0) \rangle$ is 
\begin{equation}
\langle n_{\tau}^\beta (0) \bar{n}_{\tau}^\gamma (0) \rangle = \sum_\ell \frac{2 \ell+1}{4 \pi} \frac{C_{\ell}^{n_{\tau}^\beta n_{\tau}^\gamma} }{C_{\ell}^{\mathrm{pCMB}}}.
\end{equation}
In particular, note that on average in this limit the optical depth bias is independent of scale $L''M''$, and small when the differential optical depth is reconstructed well. 

For the map space estimator, we have:
\begin{equation}
\hat{v}_I^\alpha = \sum_{\beta = 0}^{N-1} \left[  \int d\Omega \ W_I (\hat{n}) \hat{\tau}^\beta (\hat{n}) \hat{\bar{\tau}}^\alpha (\hat{n}) \right]^{-1} \int d\Omega \ W_I (\hat{n}) \hat{\tau}^\beta (\hat{n}) \bar{\Theta} (\hat{n}).
\end{equation}
Computing the estimator mean:
\begin{equation}
\langle \hat{v}_I^\alpha \rangle = \sum_{\gamma = 0}^{N-1} R^{\alpha \gamma}_I v_I^\gamma + \hat{\beta}_I^{\alpha}
\end{equation}
where
\begin{equation}
R^{\alpha \gamma}_I \equiv \sum_{\beta = 0}^{N-1} \left[  \int d\Omega \ W_I (\hat{n}) \hat{\tau}^\beta (\hat{n}) \hat{\bar{\tau}}^\alpha (\hat{n}) \right]^{-1} \int d\Omega \ W_I (\hat{n}) \hat{\tau}^\beta (\hat{n}) \bar{\tau}^\gamma (\hat{n})
\end{equation}
and
\begin{eqnarray}
\hat{\beta}_I^\alpha &=& \sum_{\beta; \gamma = 0}^{N-1}  \left[  \int d\Omega \ W_I (\hat{n}) \hat{\tau}^\beta (\hat{n}) \hat{\bar{\tau}}^\alpha (\hat{n}) \right]^{-1} \int d\Omega \ W_I (\hat{n}) \hat{\tau}^\beta (\hat{n}) \bar{\tau}^\gamma (\hat{n}) \Delta v^\gamma_I (\hat{n}) \nonumber \\ 
&+&\sum_{\beta=0}^{N-1} \sum_{c=N}^\infty  \left[  \int d\Omega \ W_I (\hat{n}) \hat{\tau}^\beta (\hat{n}) \hat{\bar{\tau}}^\alpha (\hat{n}) \right]^{-1} \int d\Omega \ W_I (\hat{n}) \tau^\beta (\hat{n}) \bar{\tau}^c (\hat{n}) v^c (\hat{n}).
\end{eqnarray}
Assuming a well-modelled optical depth, as in Eq.~\eqref{eq:tausnmodel}, a similar derivation as for the harmonic estimator yields:
\begin{equation}
\langle R^{\alpha \gamma}_I \rangle \simeq \delta_{\alpha \gamma} - \sum_{\beta=0}^{N-1} \langle \tau^\alpha (0) \bar{\tau}^\beta (0)\rangle^{-1} \langle n_{\tau}^\beta (0) \bar{n}_{\tau}^\gamma (0) \rangle.
\end{equation}
So, the optical depth bias in this limit is independent of the pixel index $I$.

\end{document}